\documentclass[USenglish]{article}

\usepackage[utf8]{inputenc}
\usepackage[small]{dgruyter}
\usepackage{lmodern}
\usepackage{amsmath}
\usepackage{hyperref}
\usepackage{microtype}
\usepackage{graphicx}
\usepackage{caption}
\usepackage{subcaption}
\usepackage{float}

\begin{document}

  \articletype{...}

  \author*[1]{Dhrubajyoti Biswas}
  \author[2]{Soumyajit Seth}
  \author[3]{Mita Bor}

  \runningauthor{D. Biswas, S. Seth,  M. Bor}
  \affil[1]{Department of Physics, Indian Institute of Technology Madras, Chennai - 600036. Email: dhrubajyoti98@physics.iitm.ac.in}
  \affil[2]{Nonlinear Dynamics Laboratory, Department of Physical Sciences, Indian Institute of Science Education and Research Kolkata, Mohanpur, West Bengal-741246. Email: ss14rs057@iiserkol.ac.in}
  \affil[3]{Department of Physics, St. Xavier's College (Autonomous), Kolkata-700016. Email: mbveronica96@protonmail.com}

  \title{A Study of the Dynamics of a new Piecewise Smooth Map}
  \runningtitle{A Study of the Dynamics of a new Piecewise Smooth Map}
  \subtitle{...}
  \abstract{In this article, we have studied a $1$D map, which is formed by combining the two well-known maps i.e. the tent and the logistic maps in the unit interval i.e. $[0,1]$. The proposed map can behave as the  piecewise smooth or non-smooth maps (depending on the behaviour of the map just before and after the border) and then the dynamics of the map has been studied using analytical tools and numerical simulations. Characterization has been done by primarily studying the Lyapunov spectra and the corresponding bifurcation diagrams. Some peculiar dynamics of this map have been shown numerically. Finally, a Simulink implementation of the proposed map has been demonstrated.}
  \communicated{Dhrubajyoti Biswas}
  \received{...}
  \accepted{...}
  \journalname{...}
  \journalyear{...}
  \journalvolume{..}
  \journalissue{..}
  \startpage{1}
  \aop
  \DOI{...}

\maketitle

\section{The Mathematical Setup} 

\begin{figure}[!h]
\centering
        \begin{subfigure}[b]{0.5\textwidth}
                \includegraphics[width=\linewidth, height=5cm]{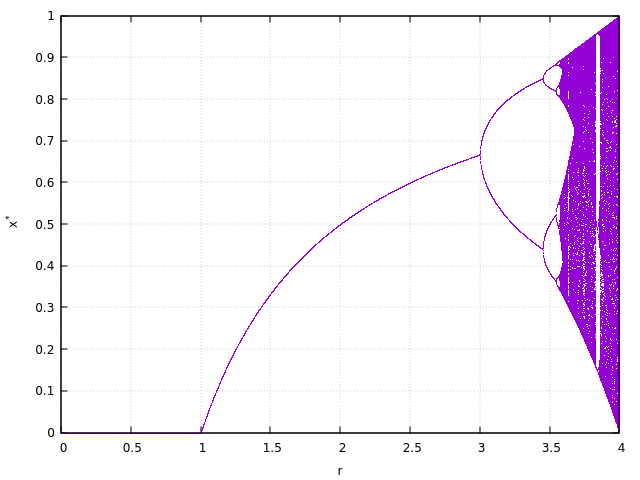}
                \caption{Bifurcation diagram of logistic map for $r\in[0,4]$. }
                \label{fig17a}
        \end{subfigure}%
        \begin{subfigure}[b]{0.5\textwidth}
                \includegraphics[width=\linewidth, height=5cm]{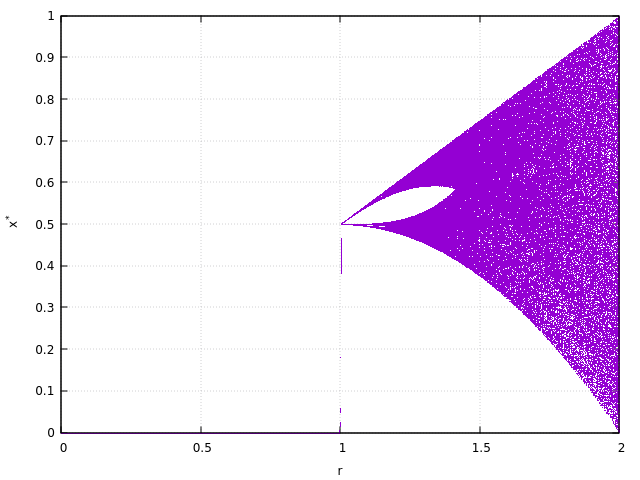}
                \caption{Bifurcation diagram of tent map for $r\in[0,2]$.}
                \label{fig17b}
        \end{subfigure}
        
\caption{Bifurcation diagram for the Logistic and Tent maps as a function of $r$.}
\label{fig17}
\end{figure}

The well known logistic and the tent maps (traditionally defined on the unit interval $[0,1]$) in the study of the chaotic dynamics \cite{strogatz2018nonlinear} are given by 
\begin{equation}
L_{r}(x)=rx(1-x)
\label{eqn1}
\end{equation}

and

\begin{equation}
T_{r}(x)=r\ min\{x,1-x\}
\label{eqn2}
\end{equation}

 respectively. The logisitc map is a quadratic continous map on the interval $[0,1]$, whereas the tent map is a piecewise linear map on the same interval. The representations ``$L_{r}$'' and ``$T_{r}$'' have been used to represent them throughout the article. The control parameters $r$ ($r\in [0,4]$ for the logistic map and $r \in [0,2]$ for the tent map) determine the dynamics exhibited by both of these.  The Tent Map gives a fixed point in the parameter interval $[0,1]$. As, the parameter is increased more, the fixed point looses it's stability and a bifurcation happens where a two piece chaotic orbit is born which gradually merges to give rise to `robust chaos' \cite{banerjee1998robust, seth2019observation} upto the parameter value $2$. As in the parameter range $[1,2]$, only the chaotic orbit exists without any periodic attractor or co-existing attractors, the chaotic orbit is `robust' in the mentioned parameter range which is shown in Figure. (\ref{fig17a}). For the case of the logistic map, it gives the period doubling bifurcation which gradually goes to chaos \cite{weisstein2001logistic}. The most interesting feature is that the periodic windows exist in-between chaotic orbits in the bifurcation parameter range, which is shown in Figure. (\ref{fig17b}). These maps have been widely studied along with their applications in population modelling, stock market profiling, encryption-decryption systems etc. in various literature over the year \cite{tsuchiya1997complete, kocarev2001logistic, habutsu1991secret, kendall1998spatial}.

\begin{figure}[t]
\centering
        \begin{subfigure}[b]{0.35\textwidth}
                \includegraphics[width=\linewidth, height=4cm]{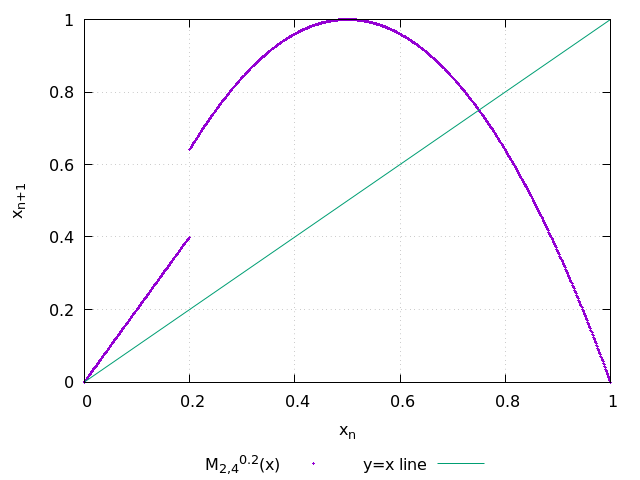}
                \caption{$\alpha=0.2$}
                \label{fig1a}
        \end{subfigure}%
        \begin{subfigure}[b]{0.35\textwidth}
                \includegraphics[width=\linewidth, height=4cm]{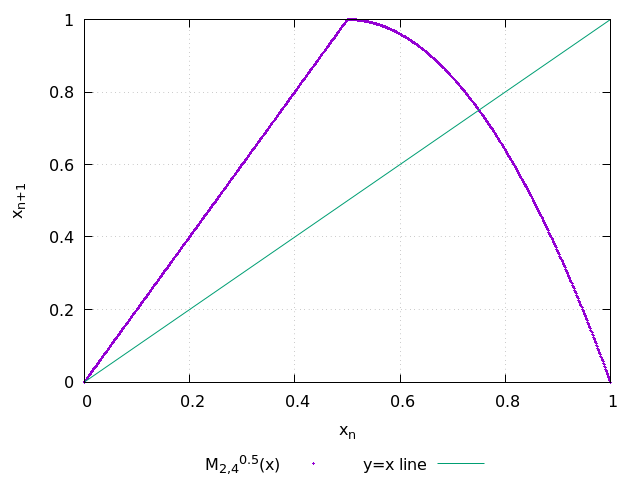}
                \caption{$\alpha=0.5$}
                \label{fig1b}
        \end{subfigure}%
        \begin{subfigure}[b]{0.35\textwidth}
                \includegraphics[width=\linewidth, height=4cm]{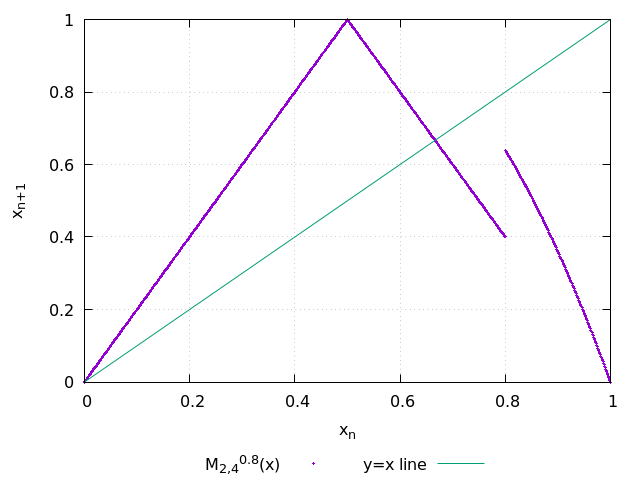}
                \caption{$\alpha=0.8$}
                \label{fig1c}
        \end{subfigure}%
\caption{Structure of $M_{2,4}^{\alpha}$ for $\alpha=0.2,0.5,0.8$}\label{fig1}
\end{figure}

This article deals with a map born out of an amalgamation of these two well known maps, aptly named as the ``Mixed Map'' (MM), denoted by ``M'', and is defined as follows:

\begin{equation}
x_{\rm n+1}=M_{r_{1}r_{2}}^{\alpha}(x_{\rm n})=\begin{cases}
T_{r_{1}}(x_{\rm n}),   \forall \ 0 \leq x_{\rm n} \leq \alpha \\
L_{r_{2}}(x_{\rm n}),   \forall \ \alpha \textless x_{\rm n} \leq 1
\end{cases}
\label{eqn3}
\end{equation}

where, $x_{\rm n}$ is the nth iteration. It is clear from the above expression that the MM is the tent map with the parameter value $r_{1}$ in the region $[0,\alpha]$ and it is the logistic map with the parameter value $r_{2}$ in the region $[\alpha,1]$. The map has a discontinuity at the point $x_{\rm n}=\alpha$.  Figure \ref{fig1} shows the structure of the map for various values of $\alpha$ for a fixed set of $(r_{1},r_{2})=(2,4)$. 

As it is clearly evident from Figure. (\ref{fig1}), there is a discontinuity at the point of transition of the map i.e. at $x_{\rm n}=\alpha$ which is known as `Border'. As the system is $1$D, the border will be a point at $x_{\rm n}= \alpha$. Depending upon the values of $\alpha$ and parameters $r_{\rm 1}$ and $r_{\rm 2}$, the map can behave as Piecewise Continuous or Piecewise Discontinuous. The map is continuous on each of the regions before and after border, but is discontinuous at border. In case of a Piecewise Discontinuous Map, there is a borderline in the Poincare section such that two arbitrarily close points on the two sides of the border land far apart at the next observation instant \cite{avrutin2006multi, jain2003border}. Otherwise, the map is known as the Piecewise Continuous Map \cite{nusse1995border, banerjee2000bifurcations}. The value of the discontinuity at border would be given by the difference in the functional values of $T_{r_{1}}$ and $L_{r_{2}}$. We can define a quantity $\delta_{r_{1,}r_{2}}(\alpha)$ as follows:

\begin{equation}
\delta_{r_{1},r_{2}}(\alpha)=\big|T_{r_{1}}(\alpha)-L_{r_{2}}(\alpha)\big|
\label{eqn4}
\end{equation} 

which can quantify the amount of discontinuity in the map at border, as a function of the value of $\alpha$. Owing to the fact that $T_{r_{1}}(x_{\rm n})$ is defined differently for $x \textgreater 0.5$ and $x \leq 0.5$, the expression of $\delta_{r_{1}r_{2}}(\alpha)$ would change for two regimes $\alpha \leq 0.5$ and $\alpha \textgreater 0.5$, which is as follows:

\begin{equation}
\delta_{r_{1},r_{2}}(\alpha)=\begin{cases} \big|r_{2}\alpha^{2}+\alpha(r_{1}-r_{2})\big|, \ \forall \alpha \leq 0.5 \\ 
\big |r_{2}\alpha^{2}-\alpha(r_{1}+r_{2})+r_{1} \big|, \ \forall \alpha \textgreater 0.5 \end{cases}
\label{eqn5}
\end{equation}

For, $\alpha\leq 0.5$, the map will have one discontinuity at $x_{\rm n} = \alpha$. Therefore the system will have one border for $\alpha \leq 0.5$. But when $\alpha \textgreater 0.5$, the system will have two borders i.e. one at $x_{\rm n} = 0.5$ and at $x_{\rm n} = \alpha$ where discontinuities will occur. Therefore, the map, which we are taking here, can have single or multiple borders depending on the value of $x_{\rm n} = \alpha$.

\begin{figure}
\includegraphics[width=\linewidth, height=5cm]{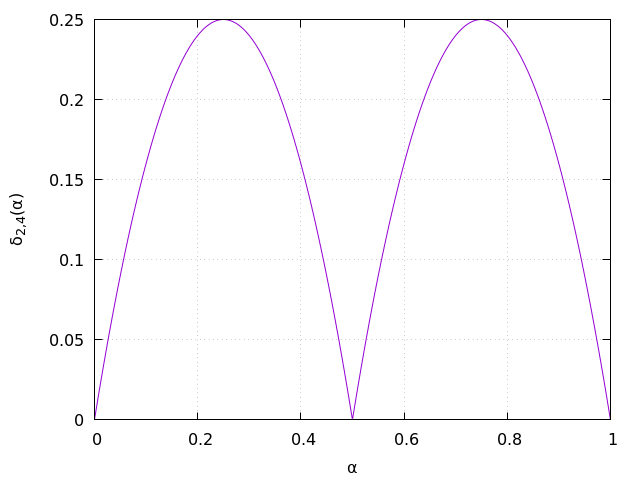}
\caption{Plot of $\delta_{\rm 2,4}(\alpha)$ vs $\alpha$. In this plot, we have shown the variation for the case of $(r_{\rm 1},r_{\rm 2})=(2,4)$. The plot agrees with the structure of the maps obtained in Figure.(\ref{fig1}) and Equations. \ref{eqn6} \& \ref{eqn7}.}
\label{fig2}
\end{figure}

Quite clearly, as seen in Figures. (\ref{fig1b}) \& (\ref{fig2}),

\begin{equation}
\delta_{2,4}(\frac{1}{2})=0
\label{eqn6}
\end{equation}

At this time the map will be piecewise continuous map with discontinuity at border $x_{\rm n} = \alpha = 0.5$.

It is quite easy to see, both mathematically and intuitively, that 

\begin{equation}
\delta_{r_{1},r_{2}}(0)=\delta_{r_{1},r_{2}}(1)=0
\label{eqn7}
\end{equation}

This is because $\alpha=1$ or $\alpha=0$ means that the existence of the Mixed Map concept is lost and it is either fully tent-like or fully logistic-like respectively.

The MM, as defined in Equation. \ref{eqn3} has three parameters which determines the dynamics of the system. While this presents a very wide range of possibilities, it becomes a hard problem to keep track of various parameters and determine their effect on the dynamics of the system when they are varied together. Thus, to simplify the problem, we define a subset of the MM, called the `Reduced Mixed Map' (RMM), in a reduced parameter space of two variables $r$ and $\alpha$, where the new parameter $r$ is related to the old ones parameters $r_{\rm 1}$ and $r_{\rm 2}$ as follows:

\begin{equation}
r_{\rm 1}=r,r_{\rm 2}=2r
\label{eqn8}
\end{equation}

\begin{figure}[!]
\centering
        \begin{subfigure}[b]{0.35\textwidth}
                \includegraphics[width=\linewidth, height=4cm]{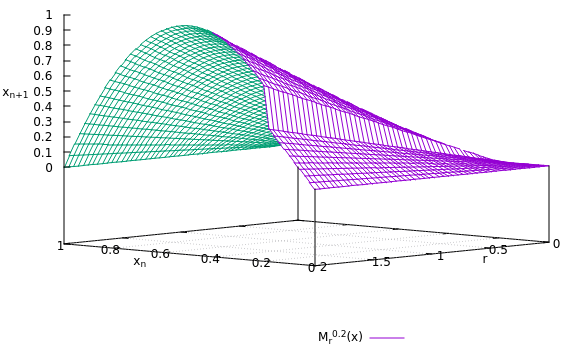}
                \caption{$\alpha=0.2$}
                \label{fig4a}
        \end{subfigure}%
        \begin{subfigure}[b]{0.35\textwidth}
                \includegraphics[width=\linewidth, height=4cm]{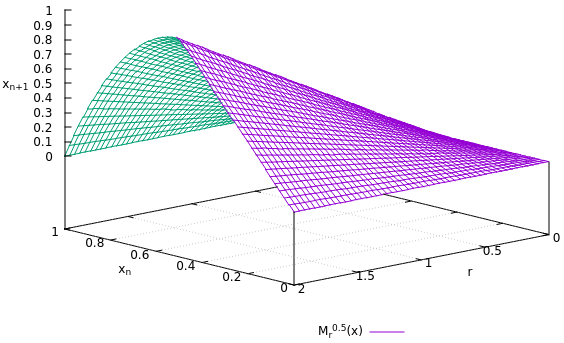}
                \caption{$\alpha=0.5$}
                \label{fig4b}
        \end{subfigure}%
        \begin{subfigure}[b]{0.35\textwidth}
                \includegraphics[width=\linewidth, height=4cm]{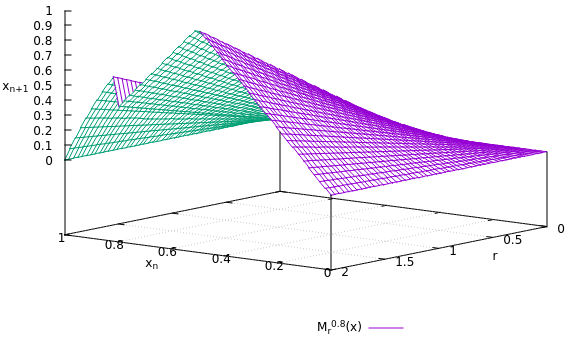}
                \caption{$\alpha=0.8$}
                \label{fig4c}
        \end{subfigure}%
\caption{Structure of $M_{r}^{\alpha}(x)$ for $\alpha=0.2,0.5,0.8$}
\label{fig3}
\end{figure}

In the definition of the RMM, the value of $\alpha$ remains the same as that of the original Mixed Map and it denotes the `Borders' of the map. The value of $r$ is bounded in the interval $[0,2]$ so as to keep the iterates of the map bounded to $[0,1]$. A convenient notation for the RMM would be to reuse the `M' notation from Equation. \ref{eqn3} as follows:

\begin{equation}
x_{\rm n+1}=M_{r}^{\alpha}(x_{\rm n})= \begin{cases}
T_{r}(x_{\rm n}),   \forall \ 0 \leq x_{\rm n} \leq \alpha \\
L_{2r}(x_{\rm n}),   \forall \ \alpha \textless x_{\rm n} \leq 1
\end{cases}
\label{eqn9}
\end{equation}

The structure of the map is shown, for various values of $\alpha$, in Figure. (\ref{fig3}). We can similarly define the amount of discontinuity at $x=\alpha$ for the RMM by plugging in Equation. \ref{eqn8} into Equation. \ref{eqn5}, which gives us:

\begin{equation}
\delta_{r}(\alpha)=\begin{cases} \big|2r\alpha^{2}-\alpha r\big|, \ \forall \alpha \leq 0.5 \\ 
\big |2r \alpha^{2} + r(1-3\alpha) \big|, \ \forall \alpha \textgreater 0.5 \end{cases}
\label{eqn10}
\end{equation}

\begin{figure}[h]
\includegraphics[width=\linewidth, height=5cm]{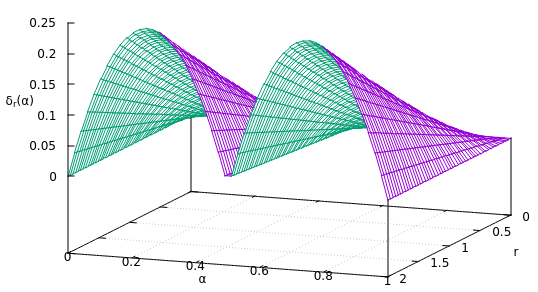}
\caption{3D plot of $\delta$ as a function of $r$ and $\alpha$. This shows how the discontinuity varies as the parameters of the system is continously changed.}
\label{fig4}
\end{figure}

Equations. \ref{eqn6} and \ref{eqn7} follows from the previous discussions as well. A 3D plot of Equation. \ref{eqn10} is shown in Figure (\ref{fig4}). Here, we have shown that the behaviour of $\delta$ if both $r$ and $\alpha$ are varied in diagram.

\section{Location and Stability of the fixed points of the RMM}

\begin{figure}[!h]
\centering
        \begin{subfigure}[b]{0.35\textwidth}
                \includegraphics[width=\linewidth, height=4cm]{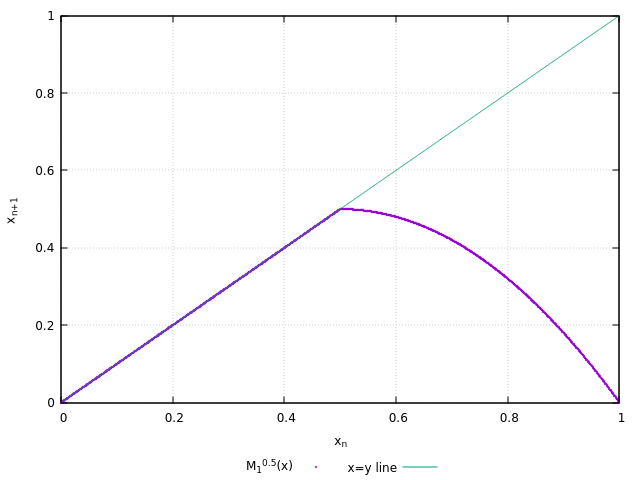}
                \caption{$\alpha=0.5$}
                \label{fig5a}
        \end{subfigure}%
        \begin{subfigure}[b]{0.35\textwidth}
                \includegraphics[width=\linewidth, height=4cm]{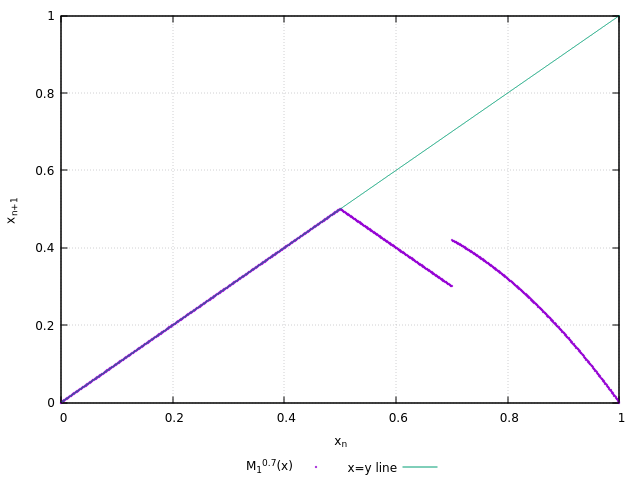}
                \caption{$\alpha=0.7$}
                \label{fig5b}
        \end{subfigure}%
        \begin{subfigure}[b]{0.35\textwidth}
                \includegraphics[width=\linewidth, height=4cm]{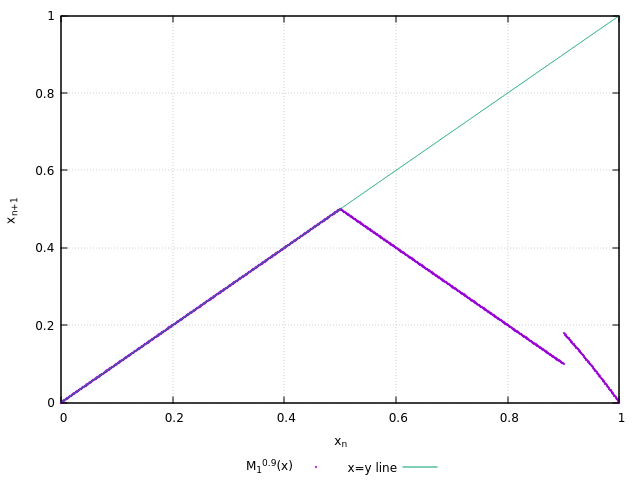}
                \caption{$\alpha=0.9$}
                \label{fig5c}
        \end{subfigure}%
        \\
        \begin{subfigure}[b]{0.35\textwidth}
                \includegraphics[width=\linewidth, height=4cm]{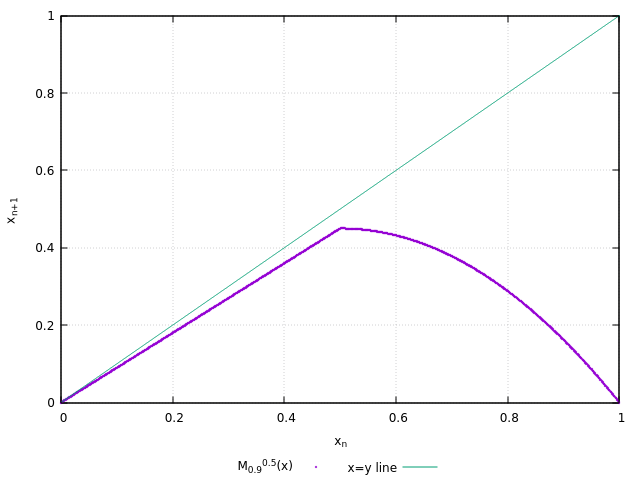}
                \caption{$\alpha=0.5$}
                \label{fig5d}
        \end{subfigure}%
        \begin{subfigure}[b]{0.35\textwidth}
                \includegraphics[width=\linewidth, height=4cm]{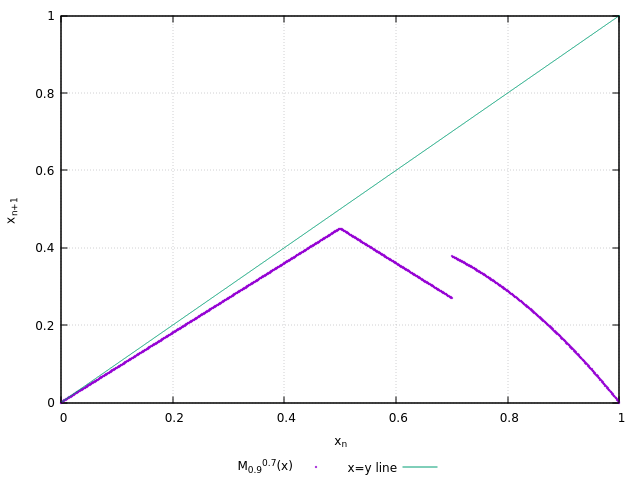}
                \caption{$\alpha=0.7$}
                \label{fig5e}
        \end{subfigure}%
       \begin{subfigure}[b]{0.35\textwidth}
                \includegraphics[width=\linewidth, height=4cm]{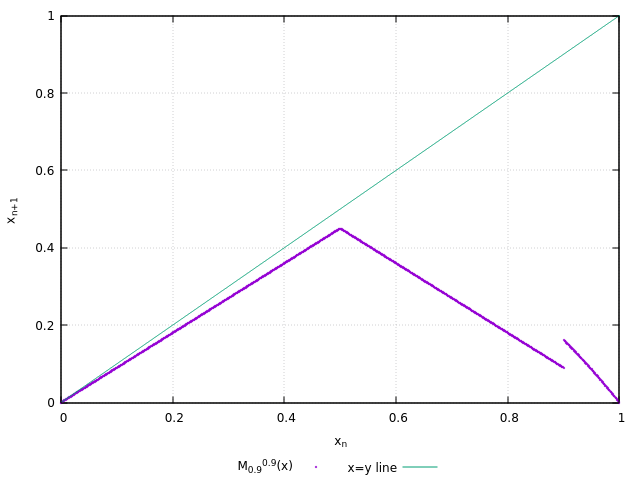}
                \caption{$\alpha=0.9$}
                \label{fig5f}
        \end{subfigure}%
       
\caption{Structure of $M_{\rm r}^{\alpha}(\rm x)$ for $\alpha=0.5,0.7,0.9$ and for $r=1,0.9$. We see that there is a line of fixed points from $x=0$ to $x=0.5$ if $r=1$ and $\alpha \geq 0.5$ in Figures. (\ref{fig5a}), (\ref{fig5b}) and (\ref{fig5c}). We also see that there are no other fixed points (other than $x_{1}^{*}=0$) if the value of $r \textless 1(=0.9)$ and $\alpha \geq 0.5$, in Figures. (\ref{fig5d}), (\ref{fig5e}) and (\ref{fig5f}).}
\label{fig5}
\end{figure}

The RMM can have atmost $3$ fixed points  whose existence, stability and expression depends on the values of $r$ and $\alpha$, with one of them always being at $x_{\rm n}^\star = 0$.
 
The existence of the fixed point $x_{\rm n}^ \star = 0$ is always guaranteed because whatever be the value of $r$ and $\alpha$, the $y=x$ line always intersects the map at the origin. The stability of the fixed point  is determined by the value of both the parameters. If $\alpha=0$, that means the map is totally logistic-like, the slope at $x_{\rm n}^\star = 0$ is given by $2r$. Then, by condition of stability,

\begin{equation}
\big| r \big| \textless \frac{1}{2}, \ \forall \ \alpha=0,\ x^{*}=0
\label{eqn11}
\end{equation}

Here, we are taking $x_{\rm n}^\star = x^ \star$.

But if the value of $1 \geq \alpha \textgreater 0$, then near to $x=0$ the map is tent-like, and thus its slope at that point is given by $r$. Thus, by condition of stability,

\begin{equation}
\big| r \big| \textless 1, \ \forall \ 1 \geq \alpha \textgreater 0, \ x^{*}=0
\label{eqn12}
\end{equation}

The other two fixed points do not always exist. If $\alpha=0$, then the map is fully logistic-like, and therefore, the only other fixed point is

\begin{equation}
x^{*}=1-\frac{1}{2r}
\label{eqn13}
\end{equation}

\begin{figure}[!h]

\centering
        \begin{subfigure}[b]{0.35\textwidth}
                \includegraphics[width=\linewidth, height=4cm]{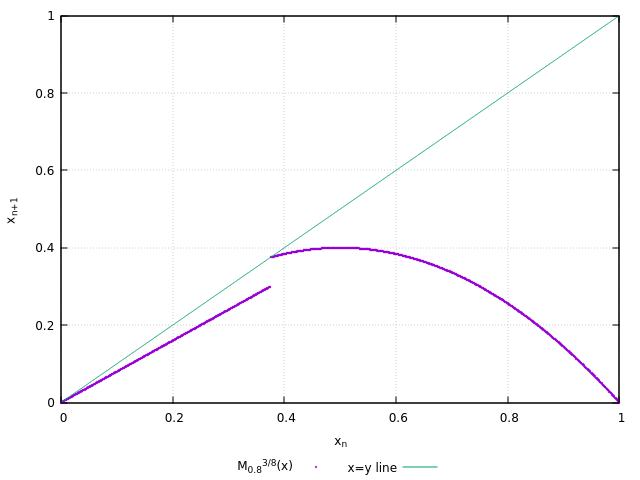}
                \caption{$\alpha=\frac{3}{8}=\alpha_{critical}$}
                \label{fig6a}
        \end{subfigure}%
        \begin{subfigure}[b]{0.35\textwidth}
                \includegraphics[width=\linewidth, height=4cm]{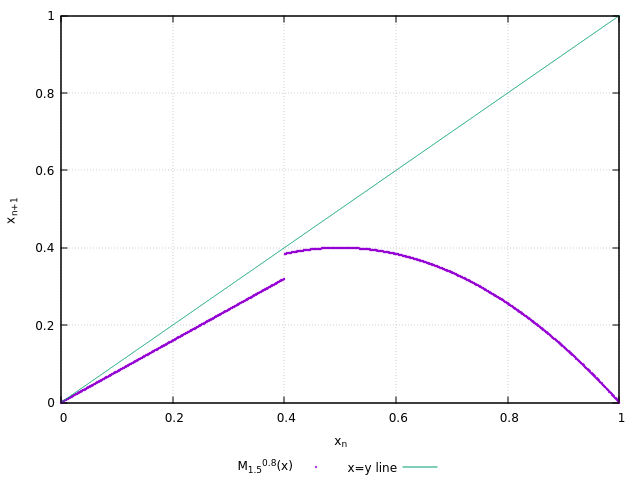}
                \caption{$\alpha=0.5 \textgreater \alpha_{critical}$}
                \label{fig6b}
        \end{subfigure}%
        \begin{subfigure}[b]{0.35\textwidth}
                \includegraphics[width=\linewidth, height=4cm]{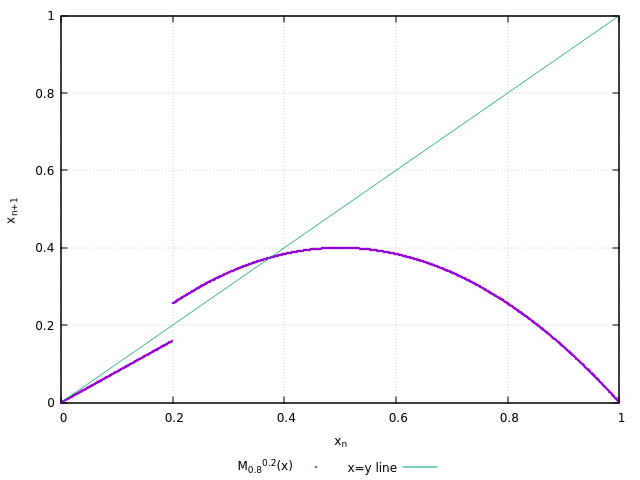}
                \caption{$\alpha=0.2 \textless \alpha_{critical}$}
                \label{fig6c}
        \end{subfigure}%
\caption{Structure of $M_{r}^{\alpha}(x)$ for $\alpha=3/8,0.5,0.2$ and for $r=0.8$. We see that there is a fixed point that appears at the value $\alpha=\alpha_{critical}$ and which remains $\forall \alpha \textless \alpha_{critical}$ and disappears for $\forall \alpha \textgreater \alpha_{critical}$}
\label{fig6}
\end{figure}

The slope at this point is given by $2-2r$, and therefore, the condition of stability is

\begin{equation}
1.5 \textgreater r \textgreater 0.5,\ \forall\ \alpha=0,\ x^{*}=1-\frac{1}{2r}
\label{eqn14}
\end{equation}

Quite clearly, the limiting case of $r=1$ results in a line of fixed points  from $x=0$ till $x=0.5$ if $\alpha \geq 0.5$. This is demonstrated in Figures. (\ref{fig5a}), (\ref{fig5b}) and (\ref{fig5c}). But if the value of $r$ is less than $1$, and $\alpha \geq 0.5$, then there is no other fixed point other than $x^{*}=0$, which is stable, having an attracting basin as that of the whole of the unit interval $[0,1]$, as demonstrated in Figures. (\ref{fig5d}), (\ref{fig5e}) and (\ref{fig5f}).

\begin{figure}[!h]
\centering
        \begin{subfigure}[b]{0.35\textwidth}
                \includegraphics[width=\linewidth, height=4cm]{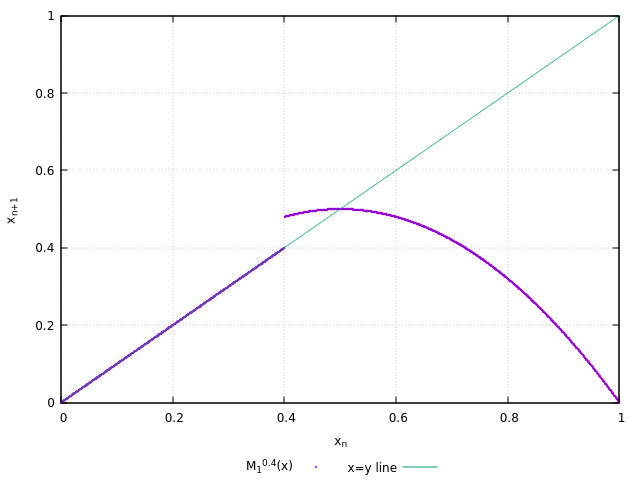}
                \caption{$\alpha=0.4$}
                \label{fig7a}
        \end{subfigure}%
        \begin{subfigure}[b]{0.35\textwidth}
                \includegraphics[width=\linewidth, height=4cm]{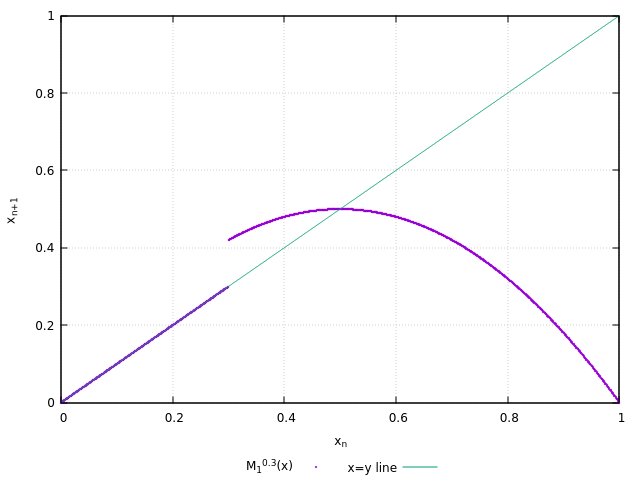}
                \caption{$\alpha=0.3$}
                \label{fig7b}
        \end{subfigure}%
        \begin{subfigure}[b]{0.35\textwidth}
                \includegraphics[width=\linewidth, height=4cm]{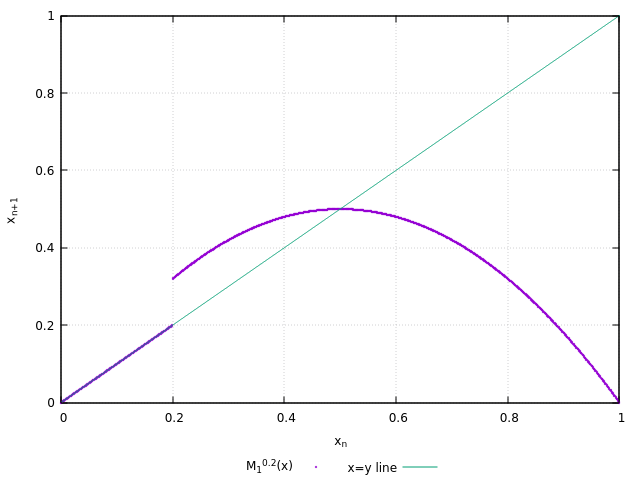}
                \caption{$\alpha=0.2$}
                \label{fig7c}
        \end{subfigure}%
\caption{Structure of $M_{r}^{\alpha}(x)$ for $\alpha=0.4,0.3,0.2$ and for $r=1$.}
\label{fig7}
\end{figure}

The case of $r \textless 1$ and $\alpha \textless 0.5$ is slightly nontrivial and there exists a fixed point other than $x^\star = 0$, given by $x^{*}=1-\frac{1}{2r}$ for $\alpha \leq (\alpha_{critical} = 1-\frac{1}{2r}) \textless 0.5$. But if $0.5 > \alpha \textgreater \alpha_{critical}$, the fixed point again vanishes leaving behind only $x^{*}=0$. This is clearly demonstrated in Figures. (\ref{fig6a}), (\ref{fig6b}) and (\ref{fig6c}). This fixed point has the same stability range as discussed in Equation. \ref{eqn14}.  Again, similar to the previous discussion, if $r=1$ and $\alpha \textless 0.5$, there again exists a fixed point given by $x^{*}=1-\frac{1}{2r}=0.5$, as shown in Figure. (\ref{fig7a}), (\ref{fig7b}) and (\ref{fig7c}), which is super-stable \cite{alligood1996chaos}. This is evident from the fact that $\big|\frac{d}{dx}M_{r}^{\alpha}(0.5)\big|=0$.

The fixed points formed on the line $x=y$ till $x=0.5$ for the case of $r=1$ and $\alpha \textgreater 0.5$ has no defined stability in the usual sense because perturbations from the fixed point neither diverge nor converge. But for initial conditions $x_{0} \textgreater 0.5$, we see the iterates converge to a fixed point rapidly (in a single step), and the fixed point is given by:

\begin{equation}
x_{x_{0}}^{*}(r)=\begin{cases}r(1-x_{0}), \ \forall \ x_{0} \leq  \alpha \\
2rx_{0}(1-x_{0}),\ \forall \ x_{o} \textgreater \alpha \end{cases}
\label{eqn15}
\end{equation}

We see that there exists a region/basin of attraction in both cases where there exists fixed points. For the case $r=1$ and $\alpha \textless 0.5$, we see that the fixed point formed at $x^{*}=0.5$ (see Figure.(\ref{fig7})) has a basin of attraction given by $x_{0} \in (\alpha,\frac{1}{2}+\frac{1}{2}\sqrt{1-2\alpha})$. For the case $r \textless 1$ and $\alpha \textless 0.5$, the stable fixed point $x^{*}=1-\frac{1}{2r}$, which only exists for $\alpha \leq \alpha_{critical}$, has a basin of attraction as same as the previous case (i.e. for $r=1$ case).

\begin{figure}[!h]
\centering
        \begin{subfigure}[b]{0.5\textwidth}
                \includegraphics[width=\linewidth, height=4cm]{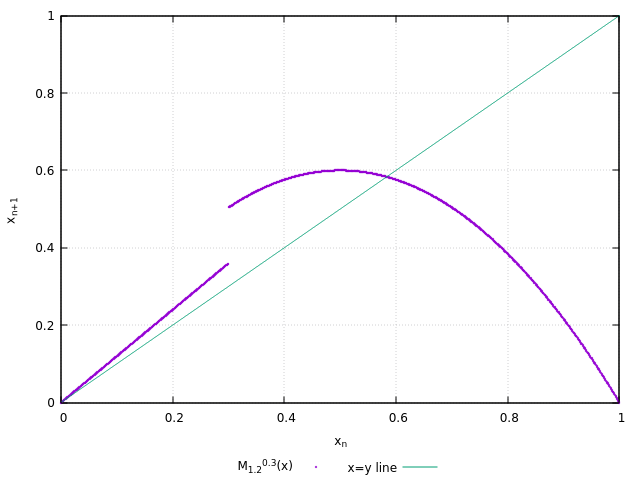}
                \caption{$\alpha=0.3$}
                \label{fig8a}
        \end{subfigure}%
        \begin{subfigure}[b]{0.5\textwidth}
                \includegraphics[width=\linewidth, height=4cm]{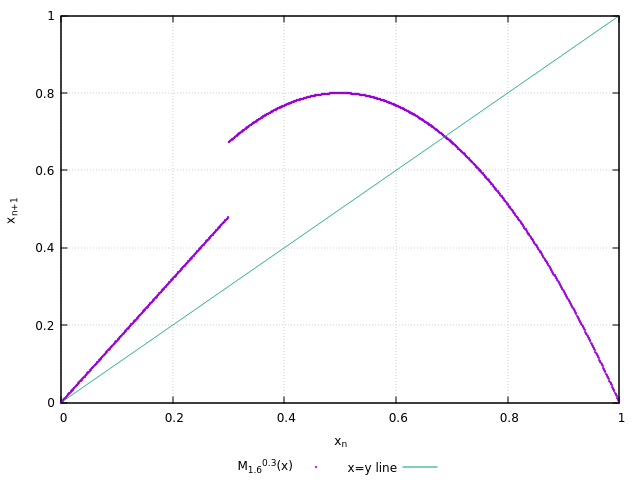}
                \caption{$\alpha=0.3$}
                \label{fig8b}
        \end{subfigure}%
        \\
        \begin{subfigure}[b]{0.5\textwidth}
                \includegraphics[width=\linewidth, height=4cm]{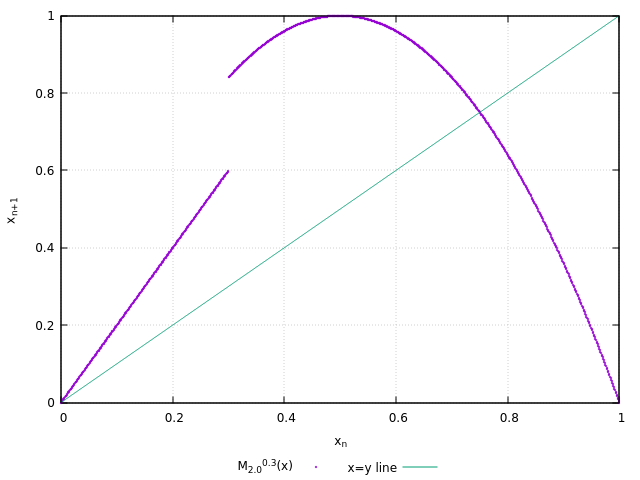}
                \caption{$\alpha=0.3$}
                \label{fig8c}
        \end{subfigure}%
        \begin{subfigure}[b]{0.5\textwidth}
                \includegraphics[width=\linewidth, height=4cm]{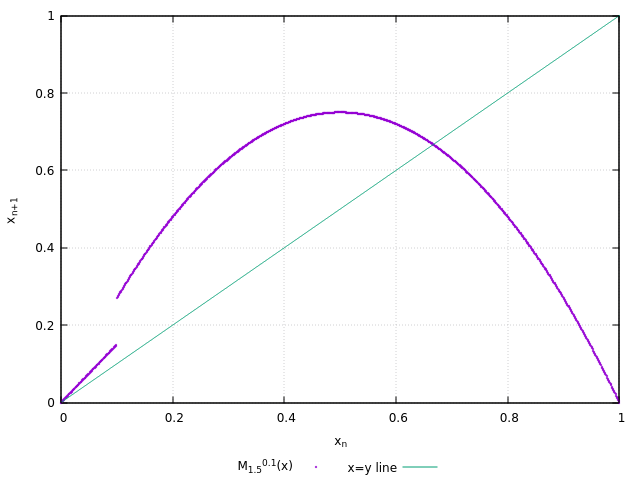}
                \caption{$\alpha=0.1$}
                \label{fig8d}
        \end{subfigure}%
        \\
        \begin{subfigure}[b]{0.5\textwidth}
                \includegraphics[width=\linewidth, height=4cm]{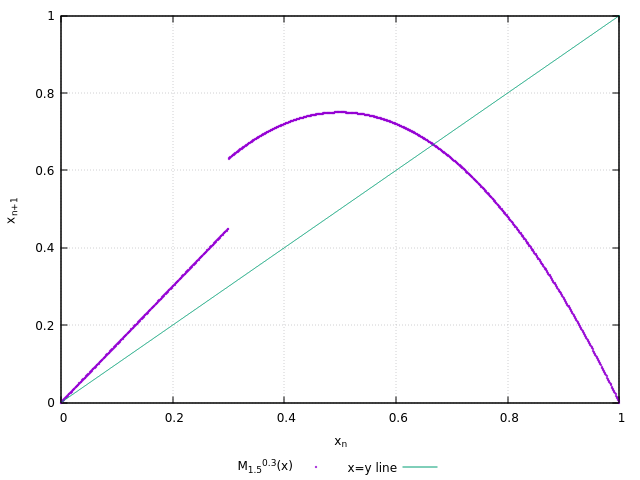}
                \caption{$\alpha=0.3$}
                \label{fig8e}
        \end{subfigure}%
        \begin{subfigure}[b]{0.5\textwidth}
                \includegraphics[width=\linewidth, height=4cm]{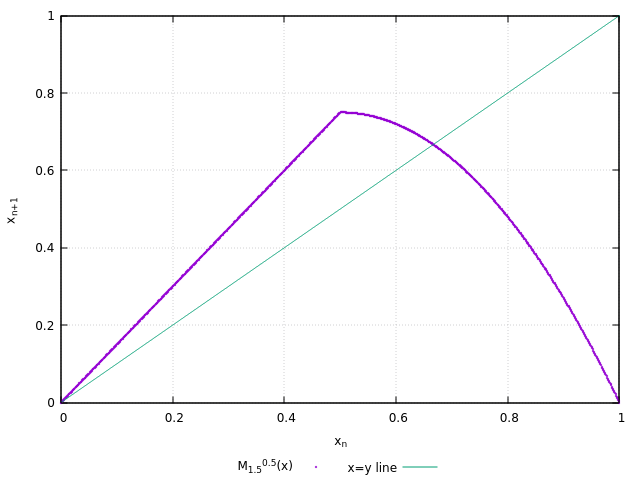}
                \caption{$\alpha=0.5$}
                \label{fig8f}
        \end{subfigure}%
\caption{Structure of $M_{r}^{\alpha}(x)$ for $\alpha=0.3$ and for $r=1.2,1.6,2.0$, in Figures. (\ref{fig8a}),(\ref{fig8b}) and (\ref{fig8c}). Structure of $M_{r}^{\alpha}(x)$ for $\alpha=0.1,0.3,0.5$ and for $r=1.5$, in Figures. (\ref{fig8d}),(\ref{fig8e}) and (\ref{fig8f}). }
\label{fig8}
\end{figure}

\begin{figure}[!h]
\centering
        \begin{subfigure}[b]{0.5\textwidth}
                \includegraphics[width=\linewidth, height=4cm]{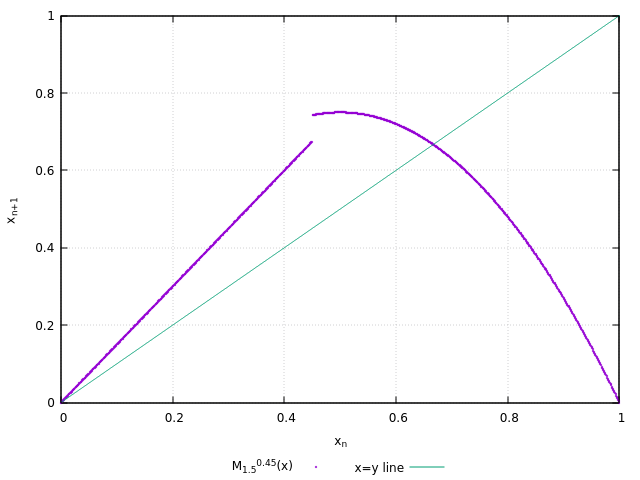}
                \caption{$\alpha \textless \alpha_{lower}$}
                \label{fig9a}
        \end{subfigure}%
        \begin{subfigure}[b]{0.5\textwidth}
                \includegraphics[width=\linewidth, height=4cm]{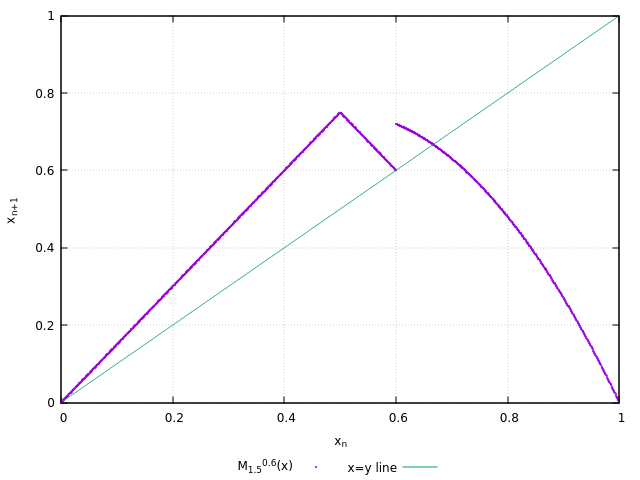}
                \caption{$\alpha=\alpha_{lower}.$ }
                \label{fig9b}
        \end{subfigure}%
        \\
        \begin{subfigure}[b]{0.5\textwidth}
                \includegraphics[width=\linewidth, height=4cm]{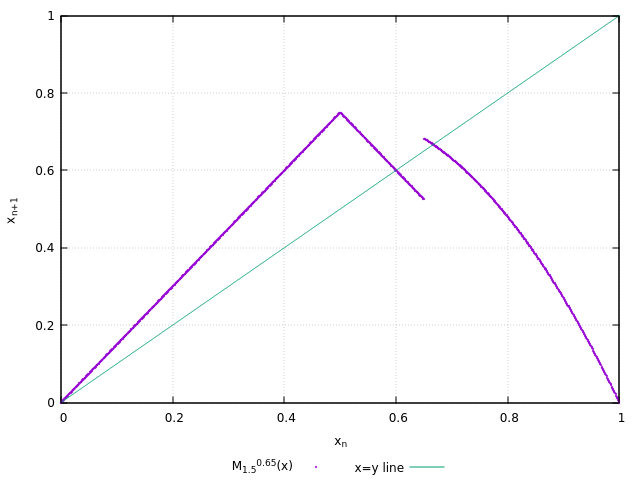}
                \caption{$\alpha \in [\alpha_{lower},\alpha_{upper}]$.}
                \label{fig9c}
        \end{subfigure}%
        \begin{subfigure}[b]{0.5\textwidth}
                \includegraphics[width=\linewidth, height=4cm]{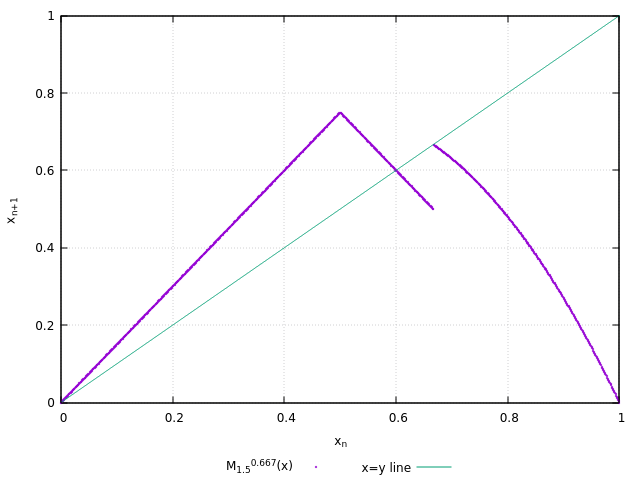}
                \caption{$\alpha=\alpha_{upper}$.}
                \label{fig9d}
        \end{subfigure}%
        \\
        \begin{subfigure}[b]{0.5\textwidth}
                \includegraphics[width=\linewidth, height=4cm]{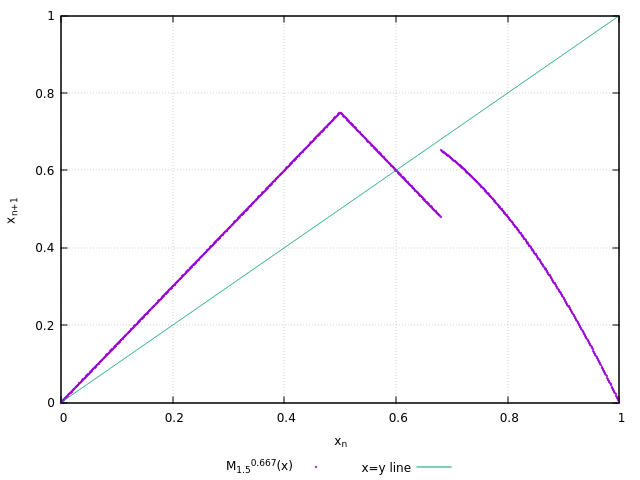}
                \caption{$\alpha \textgreater \alpha_{upper}$}
                \label{fig9e}
        \end{subfigure}%
        \begin{subfigure}[b]{0.5\textwidth}
                \includegraphics[width=\linewidth, height=4cm]{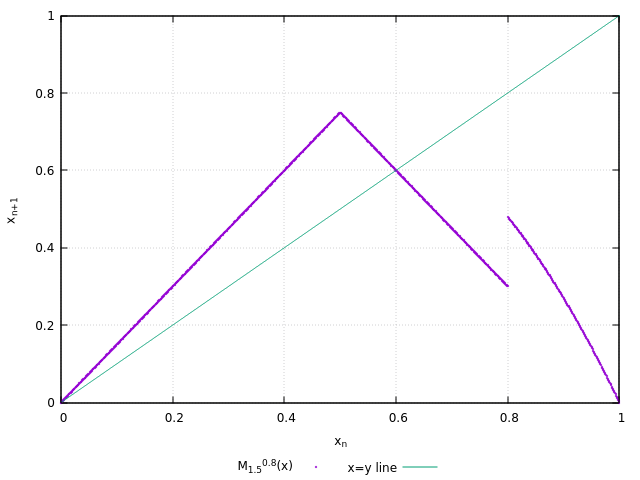}
                \caption{$\alpha \textgreater \textgreater \alpha_{upper}$}
                \label{fig9f}
        \end{subfigure}%
\caption{Structure of $M_{r}^{\alpha}(x)$ for $\alpha \textgreater 0.5$  and for $r=1.5$. Figure. (\ref{fig9a}) shows the case $\alpha \textless \alpha_{lower}$, where the fixed point is the characteristic logistic map fixed point (apart from $x^{*}=0$). Figure. (\ref{fig9b}) shows the birth of the third fixed point. Figure. (\ref{fig9c}) shows the situation where all three fixed points exist. Figure. (\ref{fig9d}) shows the situation for which the a fixed point vanishes - the fixed point which vanished is the one which existed before the third one was created in Figure. (\ref{fig9b}). Figures. (\ref{fig9e}) and (\ref{fig9f}) shows the case where only two fixed points remain (i.e. the trivial fixed point at $x^{*}=0$ and the characteristic fixed point of the tent map).}
\label{fig9}
\end{figure}

The last set of non-trivial fixed points are generated when $r \textgreater 1$. For the case $\alpha \leq 0.5$, the map again has only two fixed points, $x^{*}=0$ (which is unstable) and $x^{*}=1-\frac{1}{2r}$, which is globally stable upto $r=1.5$ (similar to Equation. \ref{eqn14}). This is described in Figures. (\ref{fig8a}) to (\ref{fig8f}). The other possibility is that $\alpha \textgreater 0.5$, which results in a peculiar case where three fixed points can exist for a certain range of $\alpha \in [\alpha_{lower},\alpha_{upper}]$. For $\alpha \textgreater \alpha_{upper}$, the fixed point is the characteristic tent map fixed point and for $\alpha \textless \alpha_{lower}$, the fixed point is the characteristic logistic map fixed point. This has been demonstrated in Figures. (\ref{fig9a}) to (\ref{fig9f}). The values of $\alpha_{lower}$ and $\alpha_{uppper}$ is given by

\begin{equation}
\alpha_{lower}(r)=\frac{r}{1+r} 
\label{eqn16}
\end{equation}
\begin{equation}
\alpha_{upper}(r)=1-\frac{1}{2r}
\label{eqn17}
\end{equation}

\begin{table}
\begin{center}
\begin{tabular}{ |c|c|c| } 
 $r\in[0,2]$ & $\alpha\in[0,1]$ & Remarks \\ 
 \hline
 $r=1$ & $\alpha \geq 0.5$ &$\rightarrow$ Line of fixed points for $x\in[0,0.5]$ \\
       & $\alpha \textless 0.5$ &$\rightarrow$ Line of fixed points for $x\in[0,\alpha]$ \\& & and the characteristic fixed point of the logistic map.  \\ 
 \hline
 $r\textless 1$ & $\alpha \textgreater \alpha_{critical}$ & $\rightarrow$Only one fixed point at $x=0$ \\
 &  $\alpha \leq \alpha_{critical}$ & $\rightarrow$ Fixed point at $x=0$ and the \\ & & characteristic fixed point of logistic map\\
 \hline
 $r \textgreater 1$ & $\alpha \in[\alpha_{lower},\alpha_{upper}]$ & $\rightarrow$ Two fixed points of characteristics tent \\ & & and logistic map along with $x=0$ (3 fixed points overall) \\
 & $\alpha \textless \alpha_{lower}$ & $\rightarrow$ Characteristic fixed point of logistic map \\ & & along with $x=0$ \\
 & $\alpha \textgreater \alpha_{upper}$ & $\rightarrow$ Characteristic fixed point of tent map \\ & &  along with $x=0$ \\
\end{tabular}
\caption{Summary of all the fixed point for various regimes.}
\label{table1}
\end{center}
\end{table}

These can be obtained easily by finding the first intersection point of the tent map and the logistic map respectively with the $y=x$ line. The fixed points themselves are the characteristic fixed points of the tent and the logistic maps and their stability has been discussed before for various values of $r$. The existance and stabilities of all the fixed points have been shown in Table \ref{table1}.

\section{Numerical Results}

\begin{figure}[!h]
\centering
        \begin{subfigure}[b]{0.5\textwidth}
                \includegraphics[width=\linewidth, height=5cm]{fig10a}
                \caption{Bifurcation diagram of logistic map for $r\in[0,4]$. }
                \label{fig10a}
        \end{subfigure}%
        \begin{subfigure}[b]{0.5\textwidth}
                \includegraphics[width=\linewidth, height=5cm]{fig10b}
                \caption{Bifurcation diagram of tent map for $r\in[0,2]$.}
                \label{fig10b}
        \end{subfigure}
        \\
        \begin{subfigure}[b]{0.5\textwidth}
                \includegraphics[width=\linewidth, height=5cm]{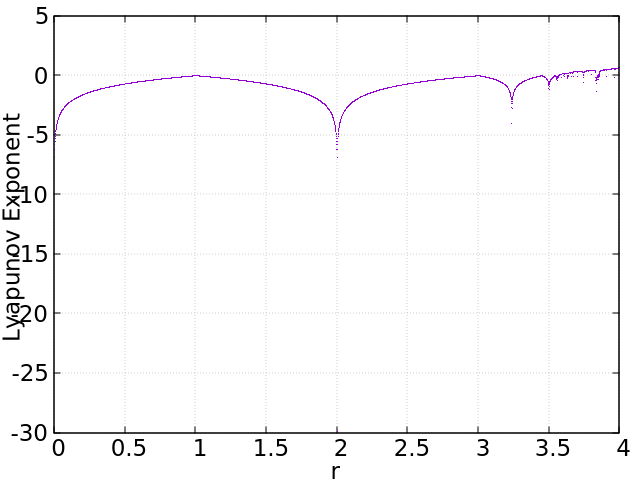}
                \caption{Lyapunov Exponent of Logistic map for $r\in[0,4]$.}
                \label{fig10c}
        \end{subfigure}%
        \begin{subfigure}[b]{0.5\textwidth}
                \includegraphics[width=\linewidth, height=5cm]{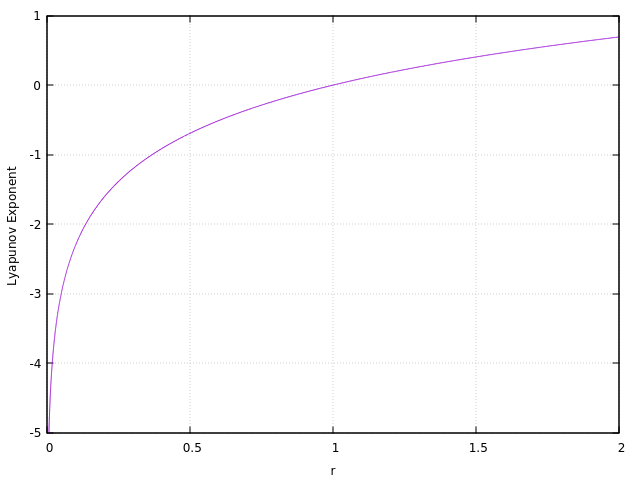}
                \caption{Lyapunov Exponent of Tent map for $r\in[0,2]$.}
                \label{fig10d}
        \end{subfigure}
\caption{Bifurcation diagram and Lyapunov Exponents for the Logistic and Tent maps as a function of $r$.}
\label{fig10}
\end{figure}

In this section, we will present the various numerical results obtained for the RMM for various values of $\alpha$ and $r$. Bifurcation diagrams form the main interest in this section because it clearly shows how the state variable of the map changes with the change of parameters.

The bifurcation diagrams of the parent maps of the RMM is shown in Figures. (\ref{fig10a}) and (\ref{fig10b}). We can see that the bifurcation diagram of the logistic map is distinctively different from that of the tent map, in terms existence of different dynamical phenomena like periodic doubling bifurcations, periodic windows in between chaos, existance of chaos without any other periodic attractors or co-existing attractors. In the case of the RMM, there are two parameters, namely, $r$ and $\alpha$, and both of them are suitable candidates for plotting the bifurcation diagrams in the various domains as discussed in the previous section. In our article, we will keep $\alpha$ as a constant, and vary $r$ and observe the bifurcation diagrams. This would be done for various different values of $\alpha$ to understand the overall behaviour of the map for various values of $\alpha$ and $r$.

\begin{figure}[!h]
\centering
        \begin{subfigure}[b]{0.5\textwidth}
                \includegraphics[width=\linewidth, height=5cm]{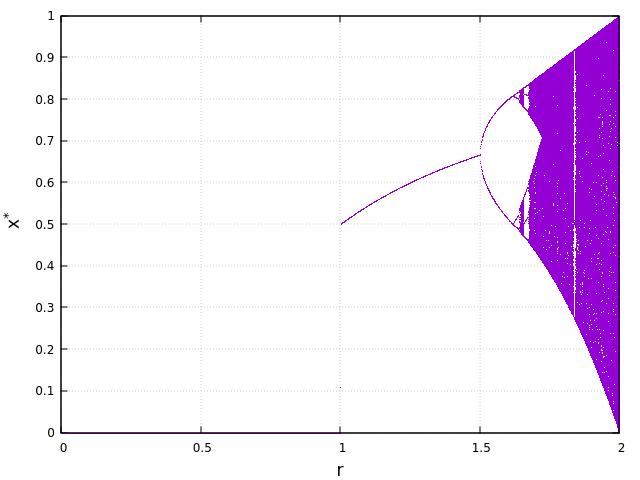}
                \caption{Bifurcation diagram of RMM for $r\in[0,2]$. }
                \label{fig11a}
        \end{subfigure}%
        \begin{subfigure}[b]{0.5\textwidth}
                \includegraphics[width=\linewidth, height=5cm]{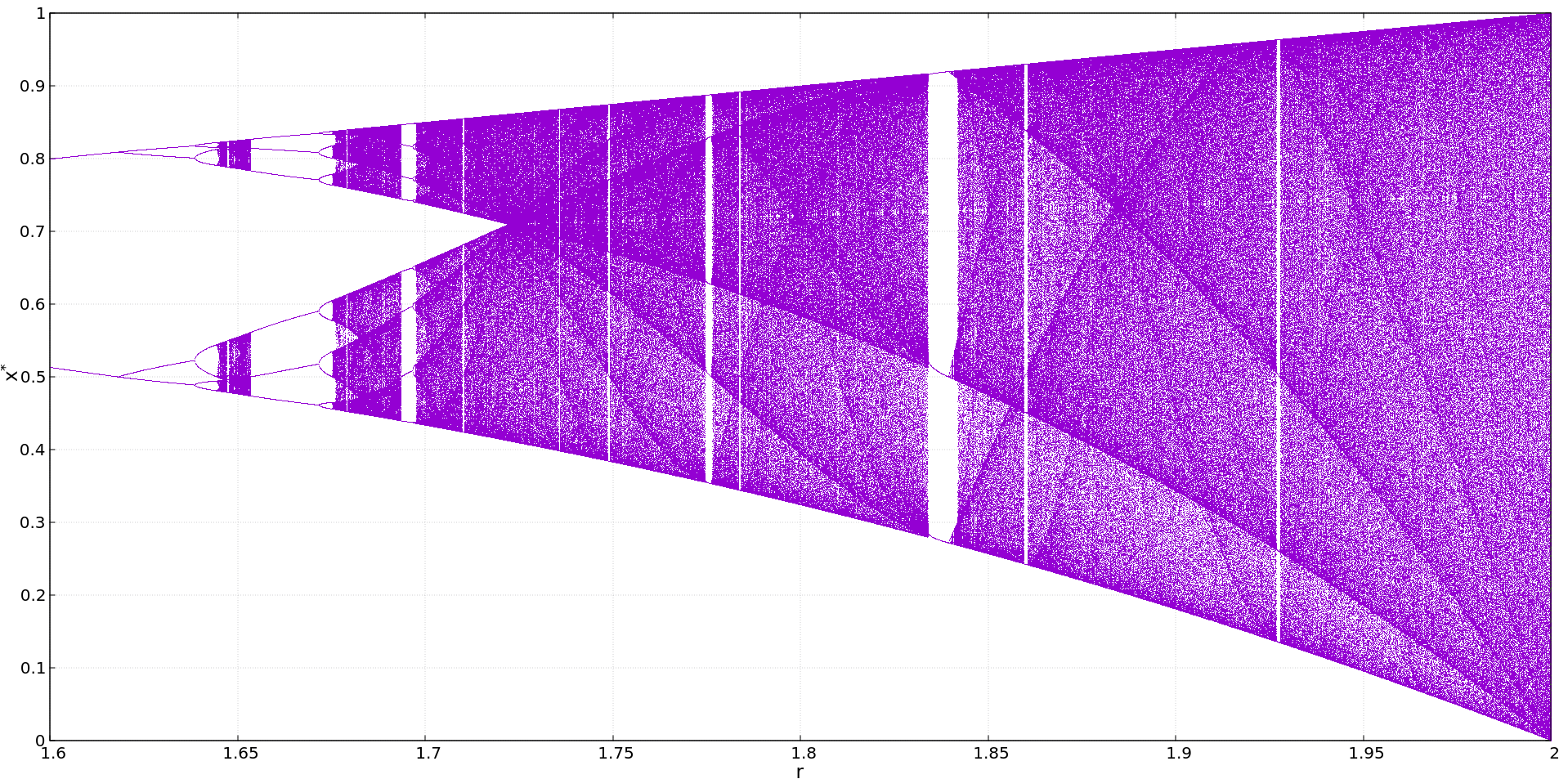}
                \caption{Bifurcation diagram of RMM for $r\in[1.6,2]$. }
                \label{fig11b}
        \end{subfigure}%
        \\
        \begin{subfigure}[b]{0.5\textwidth}
                \includegraphics[width=\linewidth, height=5cm]{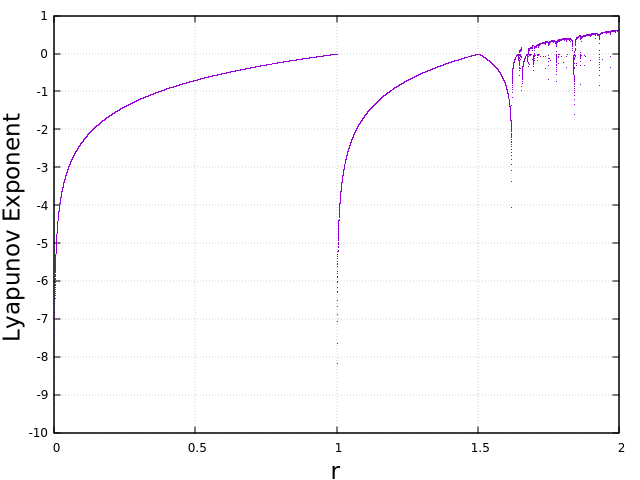}
                \caption{Lyapunov Exponent of RMM for $r\in[0,2]$. }
                \label{fig11c}
        \end{subfigure}%
        \begin{subfigure}[b]{0.5\textwidth}
                \includegraphics[width=\linewidth, height=5cm]{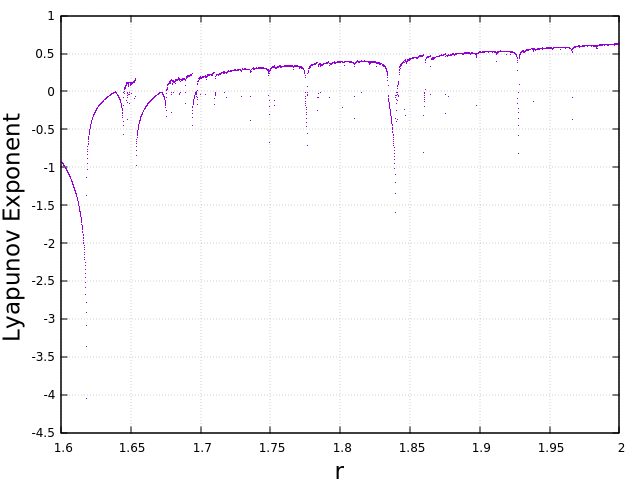}
                \caption{Lyapunov Exponent of RMM for $r\in[1.6,2]$. }
                \label{fig11d}
        \end{subfigure}%
\caption{Bifurcation diagram and Lyapunov Exponents for the RMM as a function of $r$ for $\alpha=0.5$.}
\label{fig11}
\end{figure}

If we keep the value of $\alpha=0.5$, i.e. the map is piecewise smooth as well as continuous at border, the bifurcation diagram and corresponding it's lyapunov exponents are shown in Figure. (\ref{fig11}). We see the bifurcation diagram, is in a sense, a mixture of that of the logistic map and the tent map's bifurcation diagrams. As seen in Figure. (\ref{fig10b}) in the case of the tent map, zero is a stable fixed point till $r \textless 1$. After that, the fixed point jumps to the nontrivial value and the bifurcation diagram has a structure similar to that of a logistic map where after a normal period doubling bifurcation periodic windows exist in between chaos. The corresponding lyapunov exponent shows the existence of the fixed points, periodic orbits and the values of which periodic windows exist in-between chaos. This behaviour is characterized by negative  lyapunov exponent for the fixed points and the periodic orbits, zero for the quasi-periodic orbits and positive values for the chaotic orbits.

\begin{figure}[!]
\centering
        \begin{subfigure}[b]{0.5\textwidth}
                \includegraphics[width=\linewidth, height=5cm]{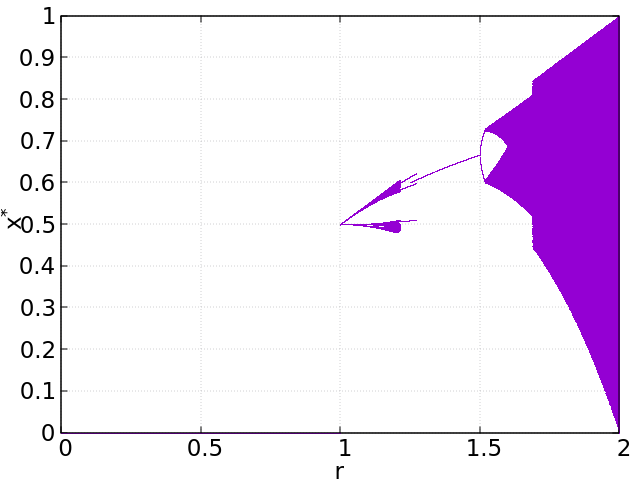}
                \caption{Bifurcation diagram of RMM for $r\in[0,2]$. }
                \label{fig12a}
        \end{subfigure}%
        \begin{subfigure}[b]{0.5\textwidth}
                \includegraphics[width=\linewidth, height=5cm]{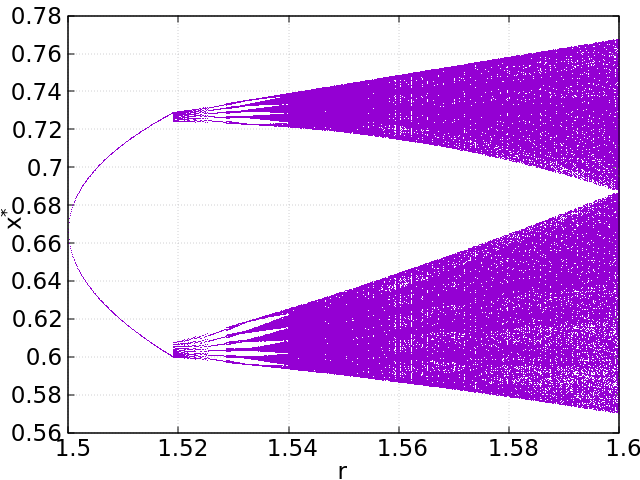}
                \caption{Bifurcation diagram of RMM for $r\in[1.5,1.6]$. }
                \label{fig12b}
        \end{subfigure}%
        \\
        \begin{subfigure}[b]{0.5\textwidth}
                \includegraphics[width=\linewidth, height=5cm]{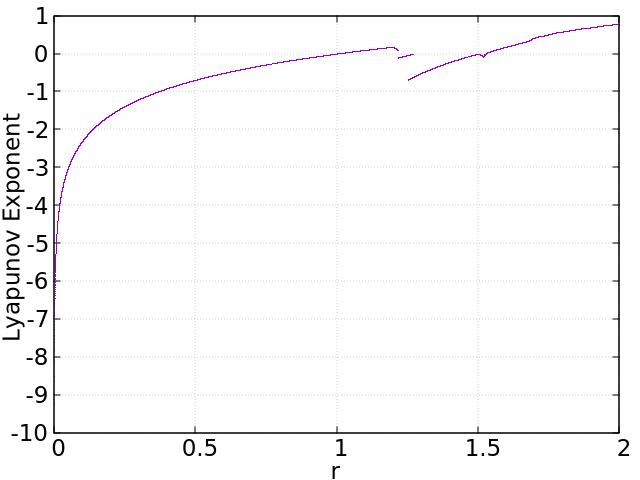}
                \caption{Lyapunov Exponent of RMM for $r\in[0,2]$. }
                \label{fig12c}
        \end{subfigure}%
        \begin{subfigure}[b]{0.5\textwidth}
                \includegraphics[width=\linewidth, height=5cm]{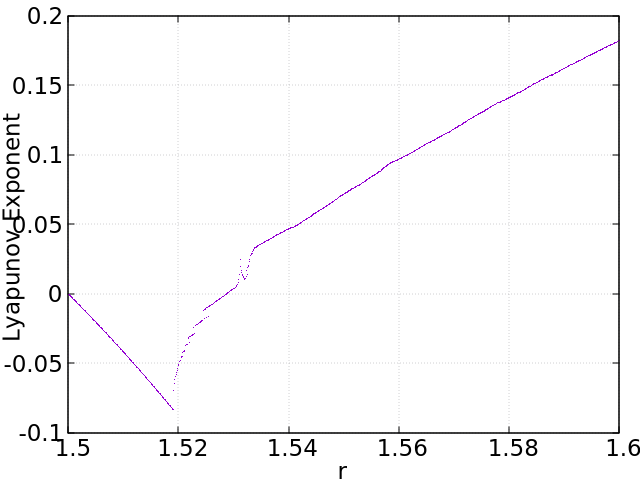}
                \caption{Lyapunov Exponent of RMM for $r\in[1.5,1.6]$. }
                \label{fig12d}
        \end{subfigure}%
        \\
        \begin{subfigure}[b]{0.5\textwidth}
                \includegraphics[width=\linewidth, height=5cm]{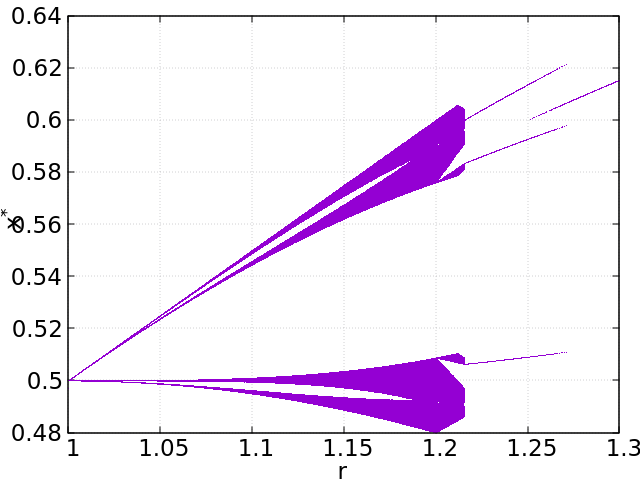}
                \caption{Bifurcation diagram of RMM for $r\in[1,1.3]$ }
                \label{fig12g}
        \end{subfigure}%
        \begin{subfigure}[b]{0.5\textwidth}
                \includegraphics[width=\linewidth, height=5cm]{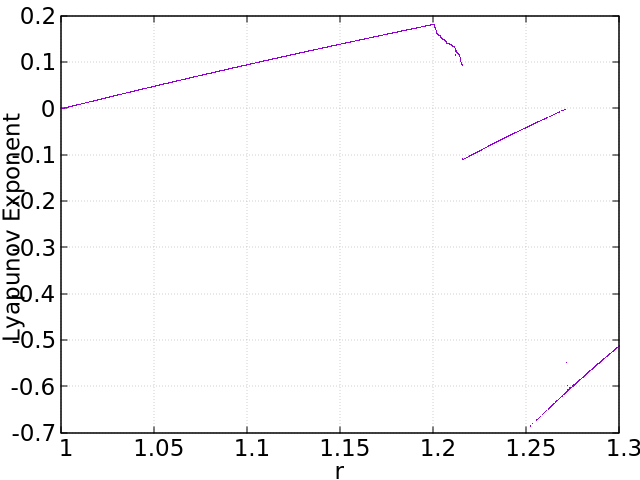}
                \caption{Lyapunov Exponent for RMM for $r\in[1,1.3]$ }
                \label{fig12h}
        \end{subfigure}%
\caption{Bifurcation diagrams and Lyapunov Exponents for the RMM for $\alpha=0.6$.}
\label{fig12}
\end{figure}

\begin{figure}[!]
\centering
        \begin{subfigure}[b]{0.5\textwidth}
                \includegraphics[width=\linewidth, height=5cm]{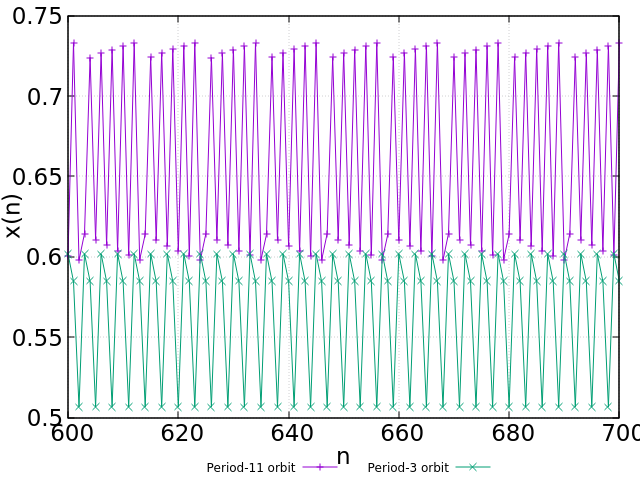}
                \caption{$r=1.528$ (period-11 orbit, purple) and $r=1.22$ (period-3 orbit, green). }
                \label{fig12e}
        \end{subfigure}%
        \begin{subfigure}[b]{0.5\textwidth}
                \includegraphics[width=\linewidth, height=5cm]{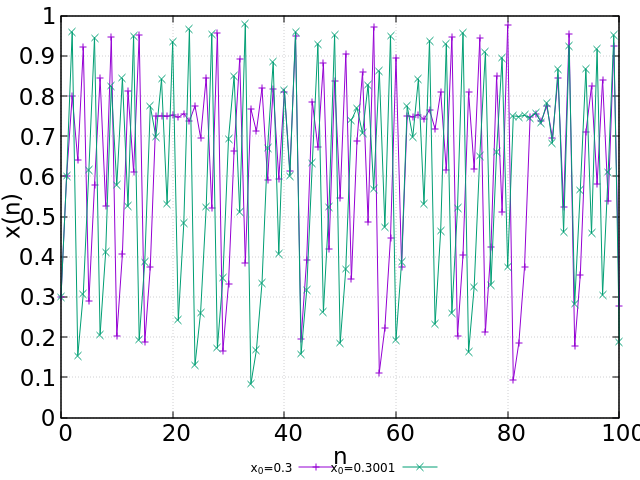}
                \caption{Time series for changes in initial conditions for $r=2$. The orbits show SIC. }
                \label{fig12f}
        \end{subfigure}%
        
\caption{Illustrations of Period $11$ and Period $3$ along with Sensitive dependence of Initial Conditions (SIC).}
\label{fig12.1}
\end{figure}

Even more interesting bifurcation diagrams emerge when we break the continuity of the map at border in case of $\alpha \neq 0.5$) i.e. the piecewise discontinuous map with single or multiple borders. If we use $\alpha=0.6$, i.e. there will be two borders, one is at $x_{\rm n} = 0.5$ and another one is at $x_{\rm n} = \alpha = 0.6$, the bifurcation diagram shows a very interesting features. There exists many interesting dynamical structure - there are period-$3$, period-$4$ upto period-$11$ orbits in this case. The time series waveforms have been shown in Figure. (\ref{fig12.1}). The period-$11$ orbit occurs near the value of $r \approx 1.528$ whereas the distinct period-$3$ orbit exists near $r \approx 1.22$ (see Figure. (\ref{fig12e})).  The period-$11$ orbit cannot be seen in case of both logistic and tent map bifurcation diagram. In Figure. (\ref{fig12a}), there exists a $2$-piece chaotic orbits just after the bifurcation near $r = 1$. After that a period-$3$ orbit emerges, which gradually gives a periodic orbit as we change the parameter further. The zoomed bifurcation diagram and corresponding it's lyapunov exponents in between $1$ to $1.3$ parameter values have been shown in Figure.(\ref{fig12g}) and Figure.(\ref{fig12h}) respectively. As the parameter $r$ is varied more, another bifurcation occurs, where a normal period doubling bifurcation gives a period-$11$ orbit, which gradudally goes to $2$- piece chaotic orbit. Also, in between $1.5$ to $2$, in the $r$ parameter range, a interior crisis \cite{banerjee1999nonlinear} happens where a chaotic orbit suddenly expands it's shape. This happens when a chaotic attractor just overlaps with the co-existing unstable chaotic orbit and the main chaotic orbit suddenly expands. Apart from all these, this map also has the normal periodic points and period-$2$ orbits as well. Figure. (\ref{fig12f}) demonstrates `sensitive dependence on initial conditions' which is a strong indicator for chaos \cite{glasner1993sensitive}.

\begin{figure}[!]
\centering
        \begin{subfigure}[b]{0.35\textwidth}
                \includegraphics[width=\linewidth, height=4cm]{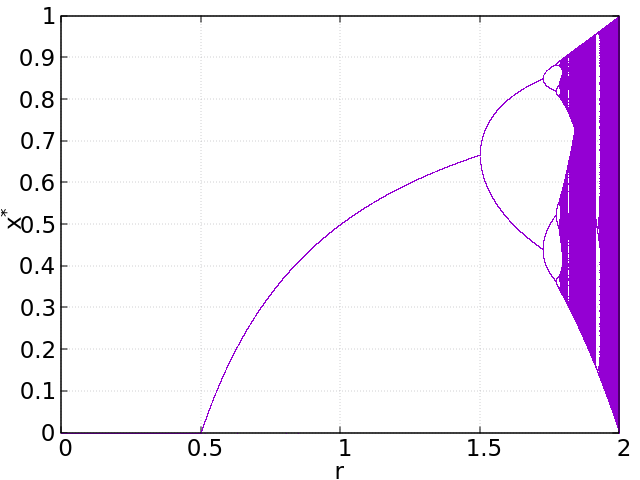}
                \caption{Bifurcation diagram of RMM for $\alpha=0.001$. }
                \label{fig13a}
        \end{subfigure}%
        \begin{subfigure}[b]{0.35\textwidth}
                \includegraphics[width=\linewidth, height=4cm]{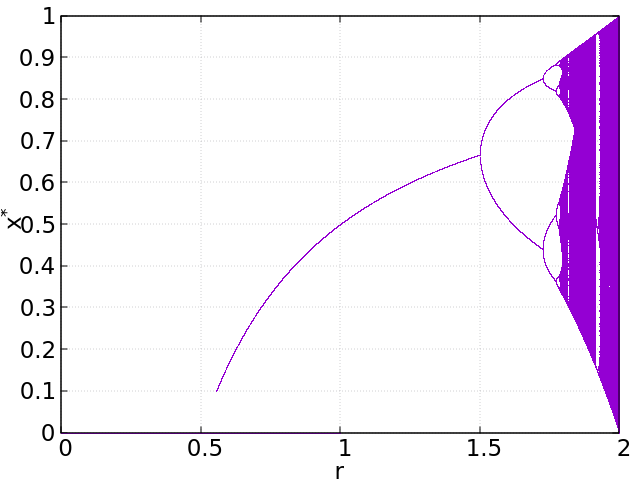}
                \caption{Bifurcation diagram of RMM for $\alpha=0.1$. }
                \label{fig13b}
        \end{subfigure}%
        \begin{subfigure}[b]{0.35\textwidth}
                \includegraphics[width=\linewidth, height=4cm]{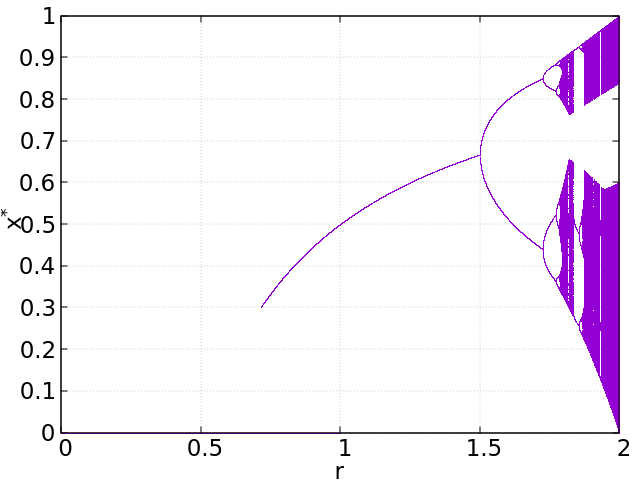}
                \caption{Bifurcation diagram of RMM for $\alpha=0.3$. }
                \label{fig13c}
        \end{subfigure}%
        \\
        \begin{subfigure}[b]{0.35\textwidth}
                \includegraphics[width=\linewidth, height=4cm]{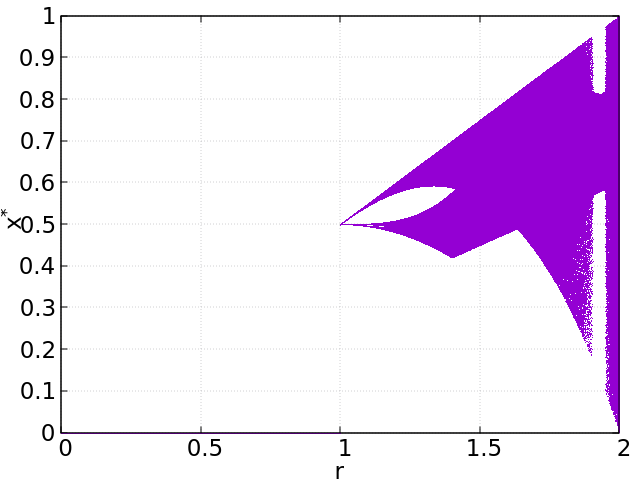}
                \caption{Bifurcation diagram of RMM for $\alpha=0.7$. }
                \label{fig13d}
        \end{subfigure}%
        \begin{subfigure}[b]{0.35\textwidth}
                \includegraphics[width=\linewidth, height=4cm]{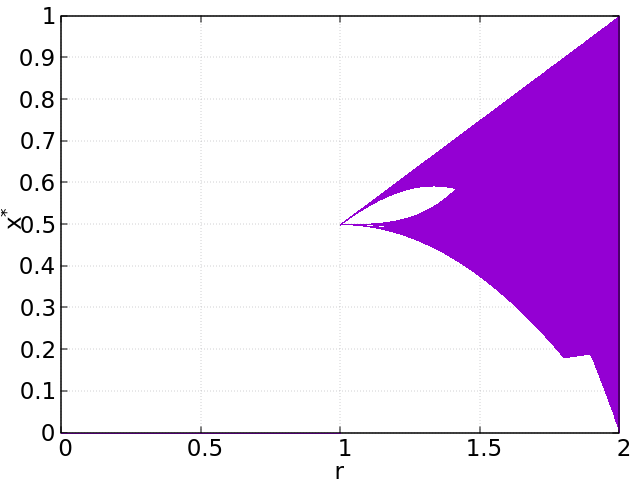}
                \caption{Bifurcation diagram of RMM for $\alpha=0.9$}
                \label{fig13e}
        \end{subfigure}%
        \begin{subfigure}[b]{0.35\textwidth}
                \includegraphics[width=\linewidth, height=4cm]{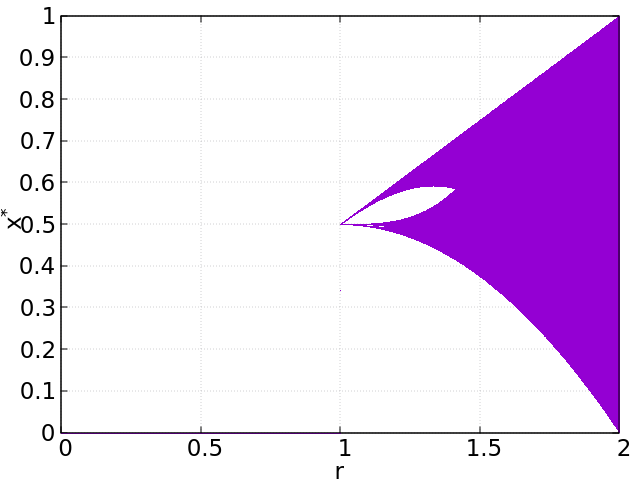}
                \caption{Bifurcation diagram of RMM for $\alpha=0.999$}
                \label{fig13f}
        \end{subfigure}%
\caption{Approach to the parent map bifurcation structures in the RMM for $\alpha \rightarrow 0$ (approaches logistic map structure, in Figures. (\ref{fig13a}), (\ref{fig13b}) and (\ref{fig13c})) and for $\alpha \rightarrow 1.0$ (approaches tent map structure, in Figures. (\ref{fig13d}), (\ref{fig13e}) and (\ref{fig13f})).}
\label{fig13}
\end{figure}

\begin{figure}[!]
\centering
		\begin{subfigure}[b]{0.35\textwidth}
                \includegraphics[width=\linewidth, height=4cm]{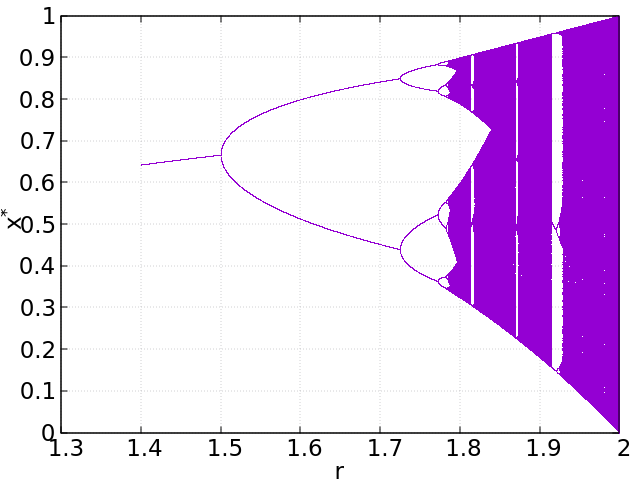}
                \caption{Bifurcation diagram of RMM for $\alpha=0.001$. }
                \label{fig13g}
        \end{subfigure}%
        \begin{subfigure}[b]{0.35\textwidth}
                \includegraphics[width=\linewidth, height=4cm]{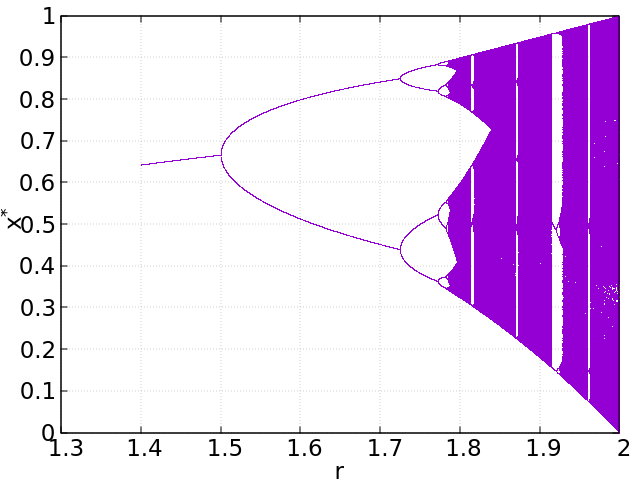}
                \caption{Bifurcation diagram of RMM for $\alpha=0.1$. }
                \label{fig13h}
        \end{subfigure}%
        \begin{subfigure}[b]{0.35\textwidth}
                \includegraphics[width=\linewidth, height=4cm]{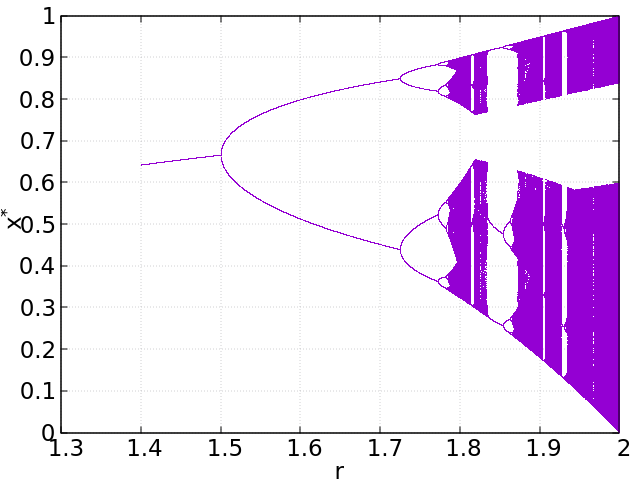}
                \caption{Bifurcation diagram of RMM for $\alpha=0.3$. }
                \label{fig13i}
        \end{subfigure}%
\caption{Zoomed Bifurcation Diagrams for the cases $\alpha=0.001$, $\alpha=0.1$ and $\alpha=0.3$.}
\label{fig13.1}
\end{figure}

\begin{figure}[!]
\centering
		\begin{subfigure}[b]{0.35\textwidth}
                \includegraphics[width=\linewidth, height=4cm]{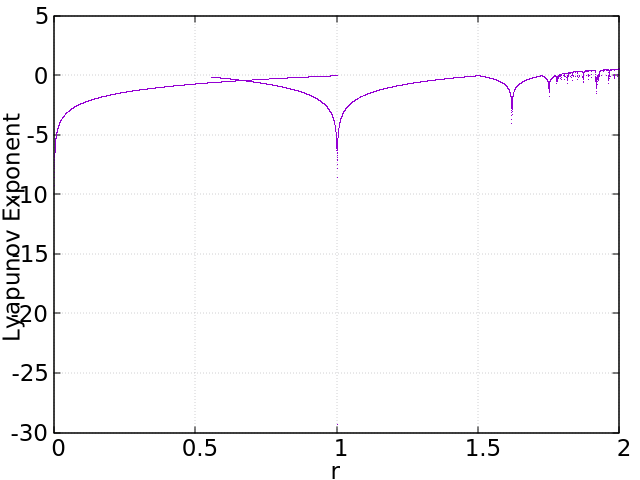}
                \caption{Lyapunov Exponent of RMM for $\alpha=0.1$. }
                \label{fig13j}
        \end{subfigure}%
        \begin{subfigure}[b]{0.35\textwidth}
                \includegraphics[width=\linewidth, height=4cm]{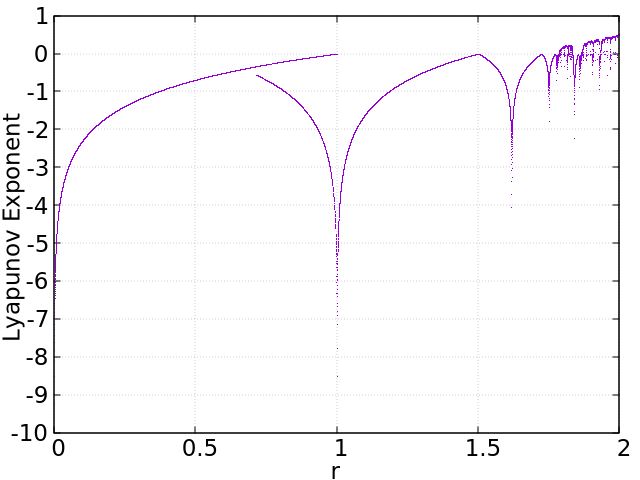}
                \caption{Lyapunov Exponent of RMM for $\alpha=0.3$.. }
                \label{fig13k}
        \end{subfigure}%
        \begin{subfigure}[b]{0.35\textwidth}
                \includegraphics[width=\linewidth, height=4cm]{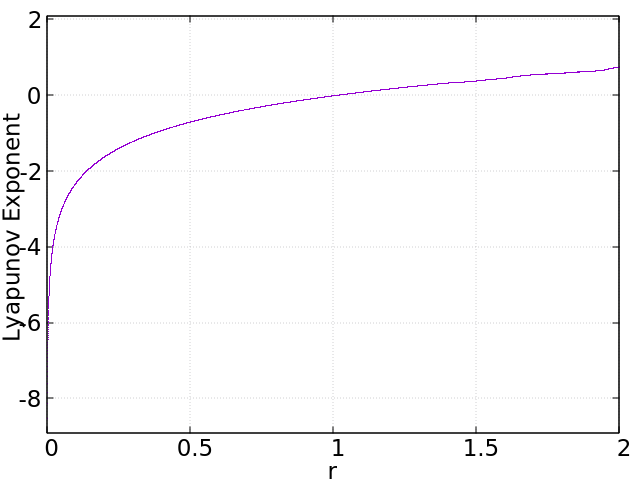}
                \caption{Lyapunov Exponent of RMM for $\alpha=0.7$.. }
                \label{fig13l}
        \end{subfigure}%
\caption{Lyapunov Exponents of RMM for $\alpha=0.1$, $\alpha=0.3$ and $\alpha=0.7$.}
\label{fig13.2}
\end{figure}

\begin{figure}[!]
\centering
		\begin{subfigure}[b]{0.35\textwidth}
                \includegraphics[width=\linewidth, height=4cm]{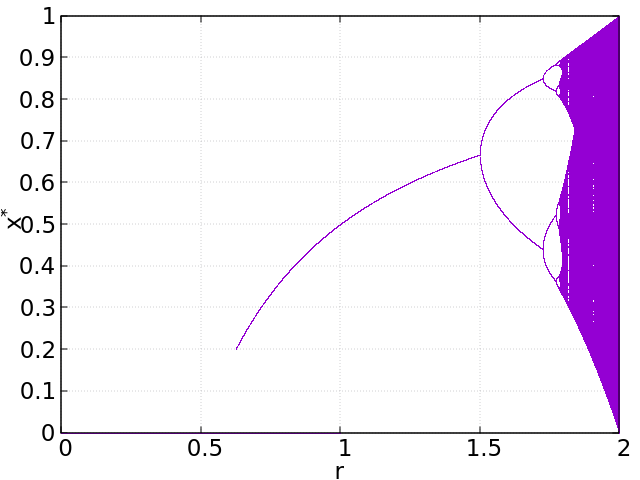}
                \caption{Bifurcation diagram of RMM for $\alpha=0.2$. }
                \label{fig14a}
        \end{subfigure}%
        \begin{subfigure}[b]{0.35\textwidth}
                \includegraphics[width=\linewidth, height=4cm]{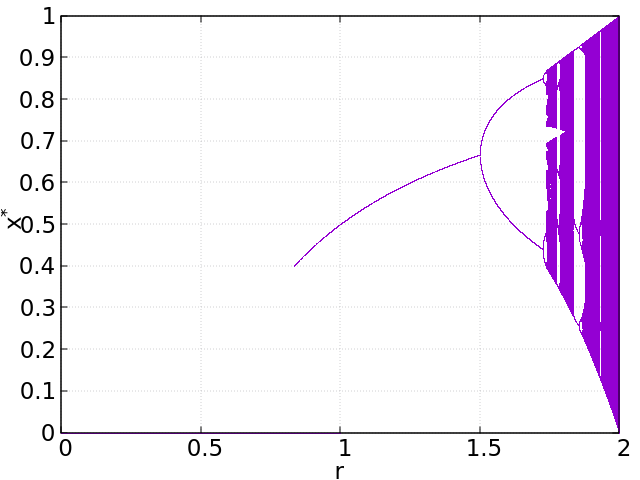}
                \caption{Bifurcation diagram of RMM for $\alpha=0.4$. }
                \label{fig14b}
        \end{subfigure}%
        \begin{subfigure}[b]{0.35\textwidth}
                \includegraphics[width=\linewidth, height=4cm]{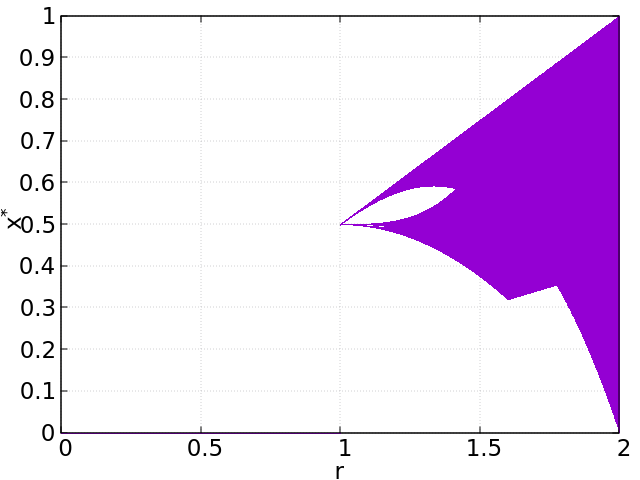}
                \caption{Bifurcation diagram of RMM for $\alpha=0.8$. }
                \label{fig14c}
        \end{subfigure}%
\caption{Bifurcation diagrams for $\alpha=0.2$,$\alpha=0.4$ and $\alpha=0.8$}
\label{fig14}
\end{figure}

\begin{figure}[!]
\centering
		\begin{subfigure}[b]{0.5\textwidth}
                \includegraphics[width=\linewidth, height=4cm]{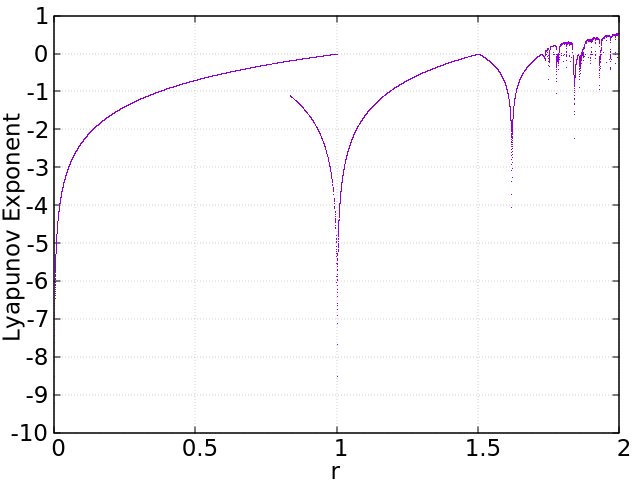}
                \caption{Lyapunov Exponent of RMM for $\alpha=0.4$. }
                \label{fig14f}
        \end{subfigure}%
        \begin{subfigure}[b]{0.5\textwidth}
                \includegraphics[width=\linewidth, height=4cm]{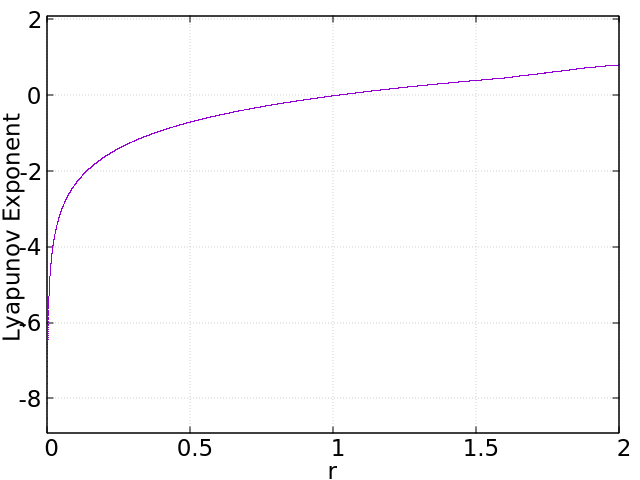}
                \caption{Lyapunov Exponent of RMM for $\alpha=0.8$.. }
                \label{fig14g}
        \end{subfigure}%
\caption{Lyapunov Exponents of RMM for $\alpha=0.4$ and $\alpha=0.8$.}
\label{fig14.1}
\end{figure}

\begin{figure}[!]
\centering
		\begin{subfigure}[b]{0.5\textwidth}
                \includegraphics[width=\linewidth, height=4cm]{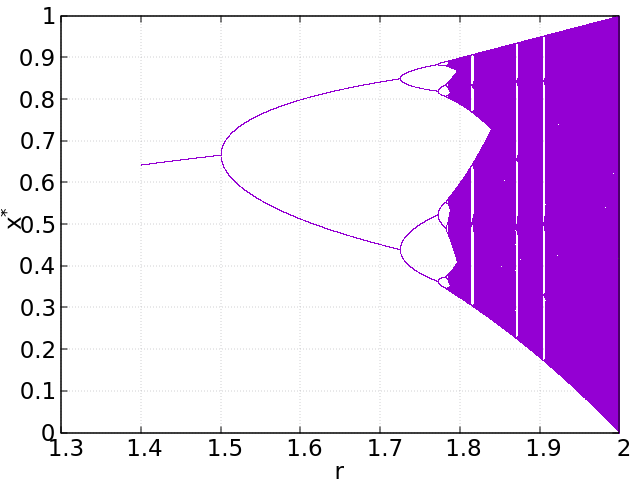}
                \caption{Bifurcation diagram of RMM for $\alpha=0.2$. }
                \label{fig14d}
        \end{subfigure}%
        \begin{subfigure}[b]{0.55\textwidth}
                \includegraphics[width=\linewidth, height=4cm]{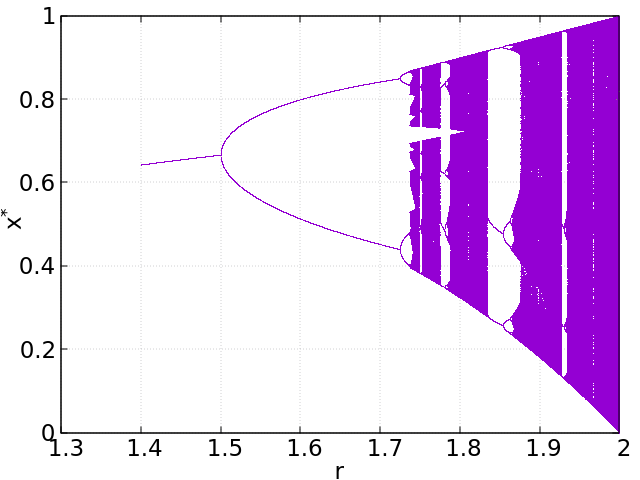}
                \caption{Bifurcation diagram of RMM for $\alpha=0.4$. }
                \label{fig14e}
        \end{subfigure}%
        
\caption{Zoomed Bifurcation Diagrams for the cases $\alpha=0.2$, $\alpha=0.4$.}
\label{fig14.2}
\end{figure}

Further changing the value of $\alpha$ gives rise to other interesting phenomena. Also, $\alpha \rightarrow 0$ leads to approach to fully logistic-like bifurcation diagram whereas $\alpha \rightarrow 1$ leads to approach to fully tent-like bifurcation diagram. These have been demonstrated in Figure. (\ref{fig13}). The zoomed versions of the bifurcation diagrams in Figures (\ref{fig13a}, \ref{fig13b}, and \ref{fig13c}) have been shown in Figure. \ref{fig13.1}. For the sake of completion, bifurcation diagrams of the case $\alpha = 0.2$, $0.4$ and $0.8$ are also shown in Figures. (\ref{fig14a}),(\ref{fig14b}) and (\ref{fig14c}). The lyapunov exponents are shown in Figures (\ref{fig13.2} and \ref{fig14.1}) respectively. In the Figures (\ref{fig13j} and \ref{fig13k}), one can see the existences of the two lyapunov exponents in the parameter values, but if one can look closely in the two diagrams, it can be said that the values of the lyapunov exponents toggle between the two lyapunov exponent values in that parameter ranges as the parameter changes gradually and as they have plotted closely, it looks like the existance of the two lyapunov exponents in the same parameter values. The bifurcation diagrams for the cases $\alpha = 0.2$ and $0.4$ have been shown in Figure (\ref{fig14.2}) for the sake of clarity in parameter range $r = 1.3$ to $2.0$.

\begin{figure}[!]
\centering
		\begin{subfigure}[b]{0.5\textwidth}
                \includegraphics[width=\linewidth, height=4cm]{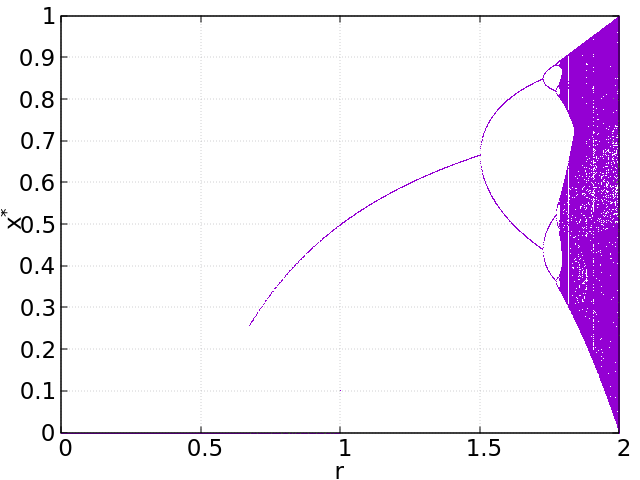}
                \caption{$\alpha=0.25$. }
                \label{fig18a}
        \end{subfigure}%
        \begin{subfigure}[b]{0.5\textwidth}
                \includegraphics[width=\linewidth, height=4cm]{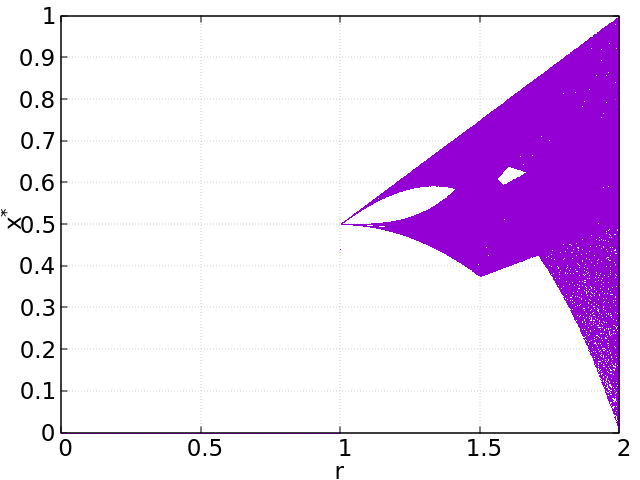}
                \caption{$\alpha=0.75$. }
                \label{fig18b}
        \end{subfigure}%
        
\caption{Bifurcation diagrams for $\alpha=0.25$ and $\alpha=0.75$. Here, $\delta=0.25$, which is same for both. }
\label{fig18}
\end{figure}

From Figure.(\ref{fig2}), it can be said that the amount of discontinuities are same in case of $\alpha = 0.25$ and $\alpha = 0.75$. But for the first case, as there is only one border which is at $\alpha = 0.25$, and for the next case, there are two borders, one is at $0.5$ and another is at $0.75$ although the amount of discontinuities are same for the two cases, the bifurcations are different due to different borders. The different bifurcation diagrams have been shown in Figure.(\ref{fig18}).

\begin{figure}[!]
\centering	
\includegraphics[width=0.7\linewidth, height=6cm]{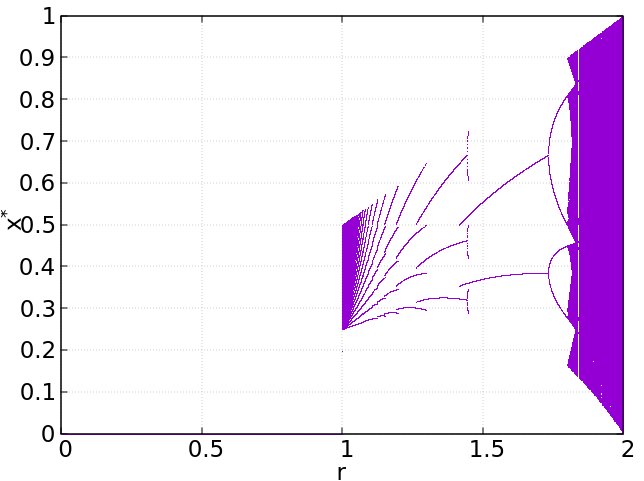}        
\caption{Bifurcation diagrams for $\alpha=0.5$ and $r_{1}=r_{2}=r$. }
\label{fig18.1}
\end{figure}

Upto this, we have taken $r_{\rm 1} = r$ and $r_{\rm 2} = 2r$. In this consideration, the map becomes continuous across the border in case of $\alpha = 0.5$. Now, we are taking $r_{\rm 1} = r_{\rm 2} = r$, which makes the map piecewise discontinuous with discontinuity at $\alpha = 0.5$.

The bifurcation in the Figure.(\ref{fig18.1}) shows a reverse period adding bifurcation upto $r = 1.6$. This phenomenon is obversed generally in case of a $1$D piecewise discontinuous map \cite{jain2003border}. As the parameter is evolved more. we get a normal period doubling bifurcations which goes to chaotic region. The periodic windows also exist in between chaos here as well.

\begin{figure}[!]
\centering
        \begin{subfigure}[b]{0.33\textwidth}
                \includegraphics[width=\linewidth, height=4cm]{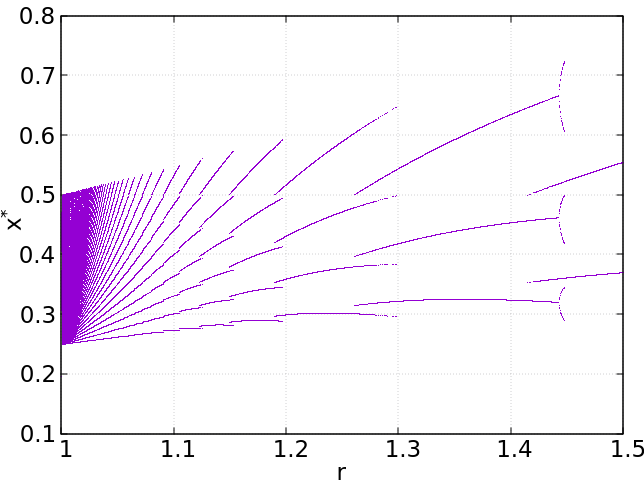}
                \caption{$\alpha=0.5$, $r\in[1,1.5]$}
                \label{fig19a}
        \end{subfigure}%
        \begin{subfigure}[b]{0.33\textwidth}
                \includegraphics[width=\linewidth, height=4cm]{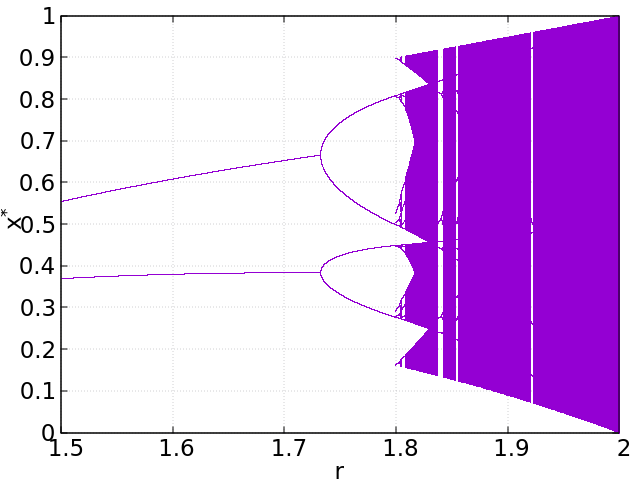}
                \caption{$\alpha=0.5$, $r\in[1.5,2]$}
                \label{fig19b}
        \end{subfigure}%
        \begin{subfigure}[b]{0.33\textwidth}
                \includegraphics[width=\linewidth, height=4cm]{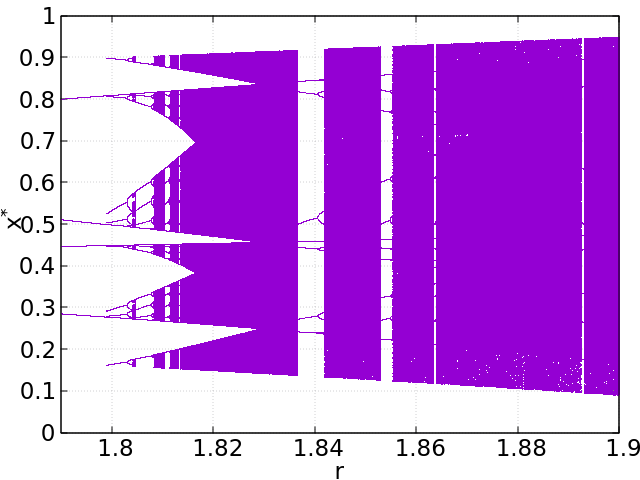}
                \caption{$\alpha=0.5$, $r\in[1.78,1.88]$}
                \label{fig19c}
        \end{subfigure}%
        \\
        \begin{subfigure}[b]{0.33\textwidth}
                \includegraphics[width=\linewidth, height=4cm]{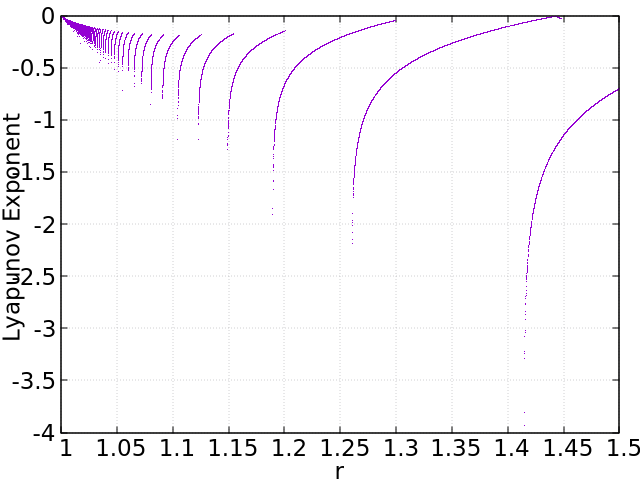}
                \caption{LE of $\alpha=0.5$, $r\in[1,1.5]$}
                \label{fig19d}
        \end{subfigure}%
        \begin{subfigure}[b]{0.33\textwidth}
                \includegraphics[width=\linewidth, height=4cm]{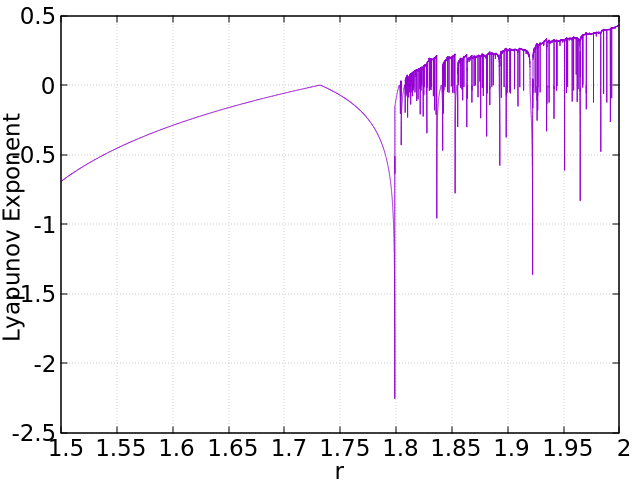}
                \caption{LE of $\alpha=0.5$, $r\in[1.5,2]$}
                \label{fig19e}
        \end{subfigure}%
        \begin{subfigure}[b]{0.33\textwidth}
                \includegraphics[width=\linewidth, height=4cm]{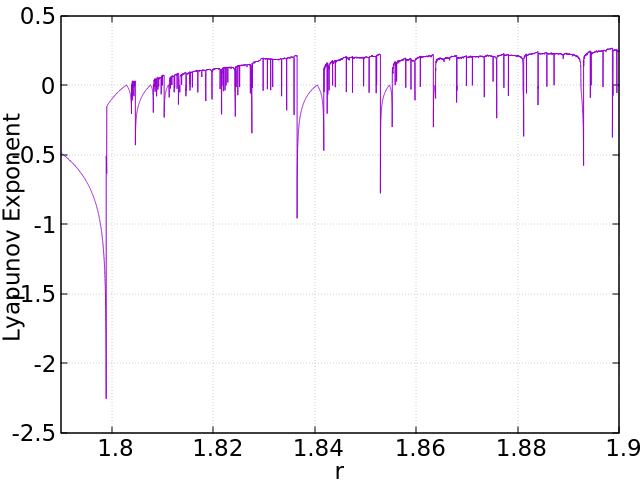}
                \caption{LE of $\alpha=0.5$,$r\in[1.78,1.8]$}
                \label{fig19f}
        \end{subfigure}%
\caption{Bifurcation diagrams and Lyapunov Exponenets for the case of $\alpha=0.5$ and $r_{1}=r_{2}=r$ of the MM.}
\label{fig19}
\end{figure}

The zoomed figures of all these bifurcations and corresponding it's lyapunov exponents have been shown in Figure.(\ref{fig19}).

\section{A Simulink based implementation of the Mixed Map}

\begin{figure}[!]
\includegraphics[width=\linewidth, height=4cm]{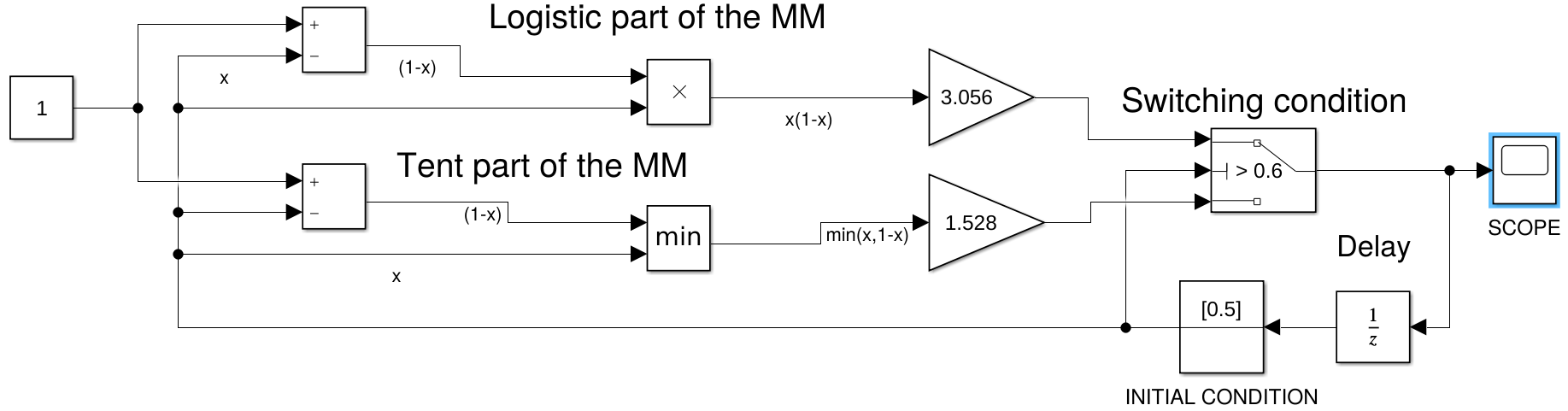}
\caption{Simulink based Implementation of the Mixed Map to generate time-series data. The initial condition, gains and the switching conditions are set arbitrarily.}
\label{fig15}
\end{figure}

In this section, we have shown a Simulink based implementation of the MM (not to be confused with the RMM, which was discussed untill now). The implementation is shown below in Figure. (\ref{fig15}).

\begin{figure}[!]
\centering
		\begin{subfigure}[b]{\textwidth}
                \includegraphics[width=\linewidth, height=5cm]{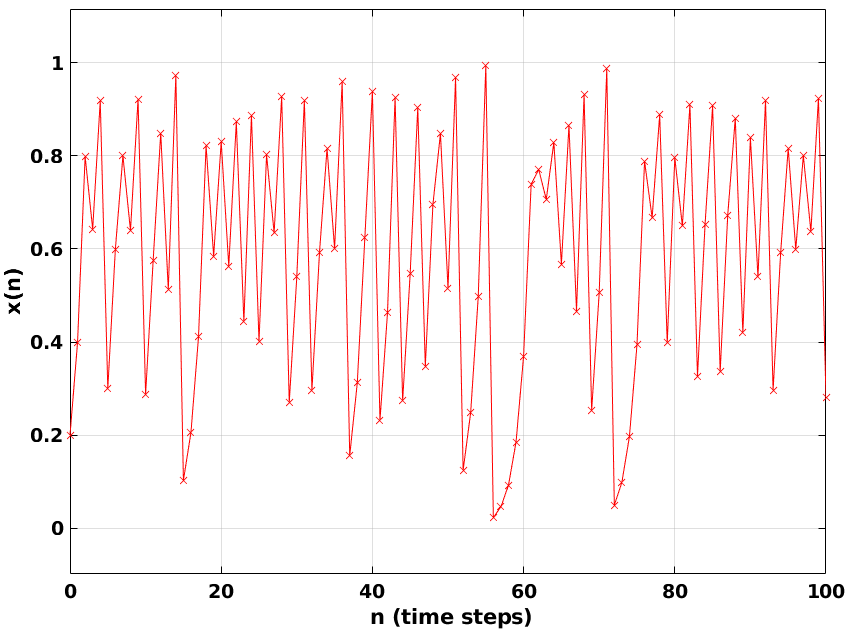}
                \caption{Time series showing aperiodic orbit. }
                \label{fig16a}
        \end{subfigure}%
        \\
        \begin{subfigure}[b]{\textwidth}
                \includegraphics[width=\linewidth, height=5cm]{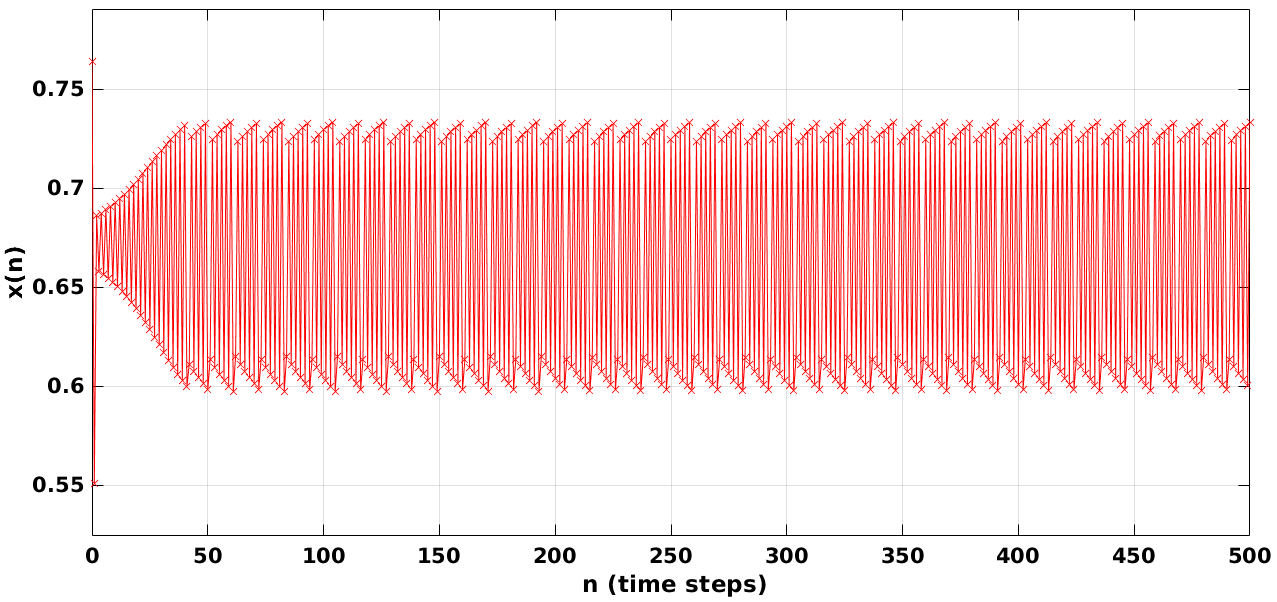}
                \caption{Time series showing period-11 orbit for $\alpha=0.6$. This has been detected previously in Figure. (\ref{fig12e}).}
                \label{fig16b}
        \end{subfigure}%
        \\
                \begin{subfigure}[b]{\textwidth}
                \includegraphics[width=\linewidth, height=5cm]{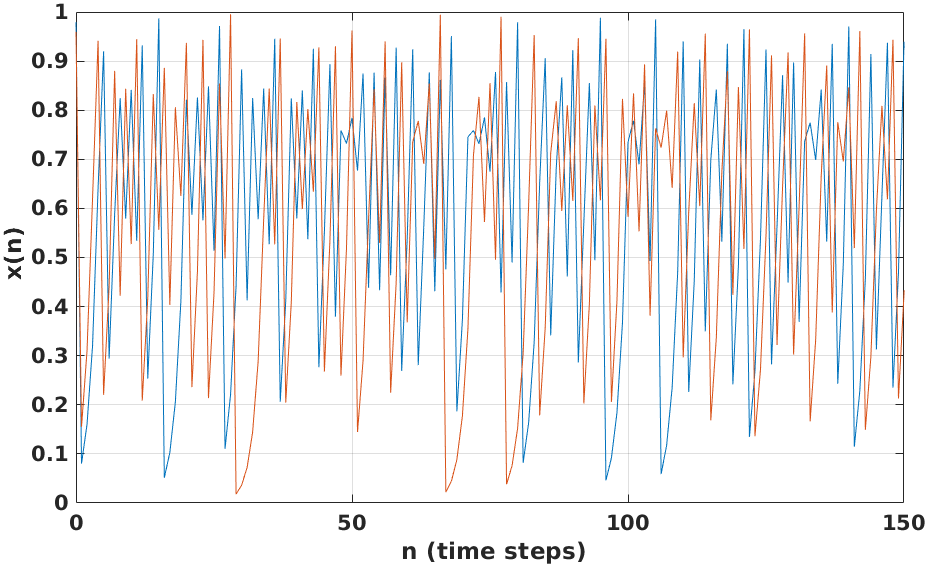}
                \caption{Time series showing SIC for two very nearby initial conditions for a pair of identical maps.}
                \label{fig16c}
        \end{subfigure}%
        \caption{Output from Simulink Implementation of MM showing Periodic orbits (Figure.(\ref{fig16a})), Non-periodic orbits (Figure.(\ref{fig16b})) and Exhibition of SIC. (Figure.(\ref{fig16c}))}
\label{fig16}
\end{figure}

The implementation as shown in Figure. (\ref{fig15}) assume the presence of various blocks (which are readily available in Simulink) like the product (multiplier), amplifier, minimum finder, controllable switch and a delay. A basic circuit diagram for implementing discrete maps is shown in \cite{suneel2006electronic}.  Following a similar method, first, the individual maps are synthesized from the available blocks. The logistic map, which is defined continuously in the interval $[0,1]$ can be easily synthesized by using a subtracter, product (multiplier) and an amplifier. The tent map, which is piecewise continuous in $[0,1]$ is implemented by using a subtracter, a minimum finder and an amplifier. The minimum finder takes two inputs and outputs the signal which is lesser of the two. The switching condition is implemented by using a controllable switch which is driven by the output of the circuit. Finally, a unit delay block is used to implement the discrete nature of our system. An initial condition block is used to set the initial condition i.e. $x_{0}$, of our system. The values of the switching parameter, the initial condition and the gains of the amplifiers (which set the control parameters $r_{1}$ and $r_{2}$ as discussed in Equation. \ref{eqn3}) can be changed manually for each simulation run. 

Some sample outputs are shown by running the simulation: Figures. (\ref{fig16a}) and (\ref{fig16b}) shows the period-11 orbit and a aperiodic orbit which were previously observed using numerical techniques. Figure. (\ref{fig16c}) exhibits Sensitive dependence on Initial Conditions (SIC) for two very nearby initial conditions for a pair of identical maps.

\section{Conclusion}
In this article, we have proposed a novel chaotic map, called the Mixed Map (MM), which is formed from the amalgamation of two well known maps i.e. logistic map and the tent map. It has two parameters $r$ and $\alpha$, which are the control parameter and the transition parameter respectively. This map was found to be both piecewise discontinous and continous as well, depending upon the values of the parameters. The amount of discontinuity was investigated and plotted. To reduce the complexity of the map, a Reduced Mixed Map (RMM) was defined for further studies. All the fixed points for the RMM for the various cases of the parameter values was found and their stability was classified. The presence of a three fixed point map was observed for certain values of the parameter $\alpha$. The results have been summarized in Table. \ref{table1}. Numerical studies were done on the proposed RMM and bifurcation diagrams, lyapunov exponents and orbits were plotted for all the interesting cases. A stable period-11 orbit was observed for $\alpha=0.6$ case. It was also shown that for $\alpha \rightarrow 0$ the bifurcation structure becomes similar to that of the logistic bifurcation structure (as because the map becomes almost totally logistic-like). For $\alpha \rightarrow 1$, the structure becomes similar to that of the tent-map bifurcation structure. For the case of $r_{\rm 1} = r_{\rm 2} =r$ and $\alpha = 0.5$ has been studied here as a special case. There, we have got a period adding cascade in reverse direction in bifurcation diagram. It can also be shown the dynamics of the map for different $\alpha$ with $r_{\rm 1} = r_{\rm 2} = r$. We can address these issues as our future work. Finally, a simple simulink implementation was presented for the proposed MM, which was used to verify some of the time series diagrams and properties we discussed in the previous sections.

\begin{acknowledgement}
DB acknowledges the HTRA scholarship provided by Indian Institute of Technology Madras, Chennai. SS acknowledges the support of DST-INSPIRE, Government of India (Ref. No: IF150667). MB acknowledges the support provided by St. Xavier's College (Autonomous), Kolkata.
\end{acknowledgement}

\bibliographystyle{plain}
\bibliography{refs}
\end{document}